\documentclass[fleqn,usenatbib]{mnras}

\usepackage{newtxtext,newtxmath}

\usepackage[T1]{fontenc}
\usepackage{ae,aecompl}

\usepackage{graphicx}	
\usepackage{amsmath}	
\usepackage{amssymb}	
\usepackage{pdflscape}

\title[A physically motivated definition for the size of galaxies]{A physically motivated definition for the size of galaxies in an era of ultra-deep imaging}

\author[Ignacio Trujillo, Nushkia Chamba \& Johan Knapen]{
Ignacio Trujillo,$^{1,2}$\thanks{E-mail: trujillo@iac.es}
Nushkia Chamba$^{1,2}$
and Johan H. Knapen$^{1,2}$ 
\\
$^{1}$Instituto de Astrof{\'i}sica de Canarias (IAC), La Laguna, 38205, Spain\\
$^{2}$Departamento de Astrof{\'i}sica, Universidad de La Laguna (ULL), E-38200, La Laguna, Spain\\
}

\date{Accepted XXX. Received YYY; in original form ZZZ}

\pubyear{2019}

\begin{document}
\label{firstpage}
\pagerange{\pageref{firstpage}--\pageref{lastpage}}
\maketitle
\begin{abstract}

Present-day multi-wavelength deep imaging surveys allow  to characterise the outskirts of galaxies with unprecedented precision. Taking advantage of this situation, we define a new physically motivated measurement of size for galaxies based on the expected location of the gas density threshold for star formation. Employing  both theoretical and observational arguments, we use the stellar mass density contour at 1 $M_{\rm \odot}$ pc$^{-2}$ as a proxy for this density threshold for star formation.  This choice makes our size definition operative. With this new  size measure, the intrinsic scatter of the global stellar mass ($M_{\rm \star}$) - size relation (explored over five orders of magnitude in stellar mass) decreases  to $\sim$0.06 dex. This value is  2.5 times smaller than the scatter measured using the effective radius ($\sim$0.15 dex) and between 1.5 and 1.8 times smaller than those using other traditional size indicators such as $R_{\rm 23.5,i}$ ($\sim$0.09 dex), the Holmberg radius $R_{\rm H}$ ($\sim$0.09 dex) and the half-mass radius $R_{\rm e,M_{\star}}$ ($\sim$0.11 dex). Moreover,  galaxies with 10$^7$ $M_{\rm \odot} <$ $M_{\star} < 10^{11}$ $M_{\rm \odot}$  increase monotonically in size following a power-law with a slope very close to 1/3,  equivalent to an average stellar mass 3D density of $\sim$4.5$\times$10$^{-3}$ $M_{\rm \odot}$ pc$^{-3}$ for  galaxies within this mass range. Galaxies with $M_{\rm \star}$$>$10$^{11}$ $M_{\rm \odot}$ show a different slope with stellar mass, which is suggestive of a larger gas density threshold for star formation at the epoch when their star formation peaks.

\end{abstract}

\begin{keywords}

galaxies: fundamental parameters - galaxies: photometry - galaxies: formation - methods: data
analysis - methods: observational - techniques: photometric 

\end{keywords}



\section{Introduction}

The sizes of  galaxies play a pivotal role in our understanding of how they form and evolve. While the size of an everyday object is  quite an intuitive concept, in the case of  galaxies
where there are no clear edges, measuring  their extent is a non-trivial task. The absence of a clear
border leads to two different ways of 
measuring the size of  galaxies in the astronomical literature. The first and today's most popular approach is
identifying the size of a galaxy as the radial distance containing half of its
light (i.e. its effective radius $R_{\rm e}$). A second and fairly common approach  to indicate galaxy size is the location of a fixed surface brightness isophote.

The effective radius has been used to characterise the size of  galaxies since  at least  the publication by \cite{1948AnAp...11..247D}. Obviously, using half
of the light of a galaxy to indicate its size is an arbitrary definition. Other
fractions of light could be and in fact have been used as well for this task, for example the
radial distance containing 90\% of the light of the galaxy ($R_{90}$) or the Petrosian and Kron radii \citep[][ for a review on how the Petrosian and Kron radii relate to the commonly used surface brightness distribution provided by the S\'ersic  (\citeyear{1968adga.book.....S}) model and how such size definitions are affected by the depth of images see \citet{2005PASA...22..118G}]{1976ApJ...209L...1P,1980ApJS...43..305K}.

One of the reasons for the popularity of $R_{\rm e}$ is its robustness against many observational issues. In particular, as the surface brightness profiles of the vast majority of  galaxies decline very rapidly (with a steepness equal to or larger than an exponential), the effective radius is barely affected by the depth of the images \citep{2001MNRAS.326..869T}. This robustness makes $R_{\rm e}$ quite appealing as a measurement for galaxy size as different authors using different datasets can reach an agreement on the size. However, despite its undeniable value, $R_{\rm e}$ is incapable of describing the global (luminous) size of galaxies  \citep[see an in-depth discussion in ][]{2019PASA...36...35G}. This limits our use of $R_{\rm e}$ as a direct measurement of galaxy size because $R_{\rm e}$ measures light concentration and strongly depends on the shape of the light profile. Consider, for example, two disc galaxies with similar appearance but with bulges of very different brightness. The global $R_{\rm e}$ of the galaxy with a prominent bulge will be significantly smaller than that of the one with a faint bulge. For this reason, and as we will
show in this work, galaxies with the same extension can have very different effective radii. This is not a minor issue and has serious consequences when one wants to address or infer the nature of galaxies \citep{chamba2020}.  In addition, the $R_{\rm e}$ of a galaxy  can vary significantly with wavelength \citep[see e.g.][]{2015MNRAS.454..806K}.

The second approach for measuring galaxy size is based on the radial location of a given isophote. The two most common size definitions are $R_{25}$  (also known as the de Vaucouleurs radius) based on the radial location of the isophote at $\mu_B$=25 mag/arcsec$^2$  and the Holmberg radius ($R_{\rm H}$) defined as the radial distance of the isophote at $\mu_B$=26.5 mag/arcsec$^2$  \citep{1958MeLuS.136....1H}.  $R_{25}$  was popularised in the famous Second Reference Catalogue of Bright Galaxies by \citet{1976RC2...C......0D}. The authors of the catalogue refer to \citet{1936MNRAS..96..588R} as the first to propose $R_{25}$ and to \citet{1960ApJ...132..306L} as the first to adopte it. These two surface brightness values correspond roughly to  10\% and 3\%  (respectively) of the brightness of the (darkest) night sky in the \textsl{B}-band in ground-based observatories.   $R_{25}$ and $R_{\rm H}$ were motivated by the typical depth of  optical images 60 years ago, and created to measure the maximum extension of  galaxies visible at that time \citep{1976RC2...C......0D}. In this sense, measuring galaxy size using such a definition was not motivated by any particular physical reason and both $R_{25}$ and $R_{\rm H}$ simply reflect the technological limitation in the 1960s. Such isophotal size definitions are not limited to the optical bands only. For example, \citet[][]{2015ApJS..219....3M} characterised the global extensions of the galaxies in the infrared Spitzer Survery of Stellar Structure in Galaxies (S$^4$G) survey \citep{2010PASP..122.1397S} using $\mu_{3.6}$=25.5 mag/arcsec$^2$ as a size indicator. In the context of exploring the scaling relations between  size, luminosity and velocity of  late-type galaxies,  \citet{2011ApJ...726...77S} and \citet{2012MNRAS.425.2741H}  found that the use of $R_{\rm 23.5,i}$ (i.e. the radial location of the isophote $\mu_i$=23.5 mag/arcsec$^2$) yields the smallest scatter in the size-luminosity relation.

In contrast to $R_{\rm e}$ and the isophotal size measures, there has also been some effort to characterise galaxy size using physically motivated parameters. An example of such a size parameter is the exponential scale length $r_{\rm d}$ which is connected with the angular momentum of dynamically stable discs \citep[see e.g.][]{1998MNRAS.295..319M,2010gfe..book.....M}. However, in practice, due to the complexity of galactic discs (which include bars, spiral arms, etc) the use of $r_{\rm d}$ has been shown to be complicated to reproduce by different authors. In fact, for the same galaxies, $r_{\rm d}$ has been measured with a scatter of $\sim$25\% \citep[see e.g.][]{1991A&A...248...57K,2004A&A...415...63M}.

All the above size measures were introduced using relatively shallow imaging surveys. More recently, however,
a revolution in the limiting depth of new astronomical imaging surveys has happened. As we will propose in this paper, image depth is no longer the limitation it once was to find a more representative and physically motivated definition for galaxy size. While the most commonly used astronomical survey, the Sloan Digital Sky Survey \citep[SDSS;][]{2003AJ....126.2081A}, reaches a comparable depth that obtained in photographic plates \citep[i.e. $\sim$26.5 mag/arcsec$^2$ in the \textsl{g}-band, which is equivalent to a 3$\sigma$ fluctuation with respect to the background of the image measured in an area of  10$\times$10 arcsec$^2$;][]{2004AJ....127..704K,2006A&A...454..759P}, surveys conducted a decade later \citep[i.e.][]{2010AJ....140..962M,2012ApJS..200....4F,2014ApJ...787L..37M,2015A&A...581A..10C,2015MNRAS.446..120D,2015ApJ...807L...2K,2016MNRAS.456.1359F,2017ApJ...834...16M} are regularly observing 2-3 mag deeper than SDSS. The current observational limit taken from ground-based telescopes is 31.5 mag/arcsec$^2$ in the \textsl{r}-band, \citep[equivalent to a 3$\sigma$ fluctuation with respect to the background of the image in an area of 10$\times$10 arcsec$^2$;][]{2016ApJ...823..123T} and a similar depth is expected to be achieved with ultra-deep surveys that are currently in operation such as the Hyper Suprime Cam Survey \citep{2018PASJ...70S...4A} and the future Large Synoptic Survey Telescope \citep[LSST;][]{2008SerAJ.176....1I} survey. Going beyond this depth has been only possible with ultra-deep imaging taken from space \citep[see e.g.][]{2019A&A...621A.133B}.

In this paper we propose a physically motivated definition to measure the size of galaxies. We suggest using the location of the gas density threshold for star formation in  galaxies as a natural size indicator, where by \textit{natural} we mean a size indicator that is connected with the intuitive concept of an edge. In other words, a size indicator that can be linked to a sharp contrast or change in the properties of the objects we explore. In practical terms,  we will show that using the radial location of the contour at a stellar mass density of 1 M$_\odot$/pc$^2$  ($R_{\rm 1}$) corresponds roughly to the location of the gas  density threshold for star formation. This definition is innately linked to the separation of the majority of stars that were born in-situ from stars that were mostly accreted throughout a galaxy's history, potentially extending its use to define the stellar halo (N. Chamba et al. in prep). In addition, as we will show below, $R_{\rm 1}$ provides  a more direct association to what an observer recognises as the total extension of a galaxy than $R_{\rm e}$. While  measuring  $R_{\rm 1}$ would have been difficult using past imaging surveys due to the required  level ($\mu_r$>26 mag/arcsec$^2$) to identify isophotes with a low mass density around 1 M$_\odot$/pc$^2$, we will see that current surveys are able to reach such depth without much difficulty.

Finally, using a physically motivated definition for measuring the size of  galaxies is not just another way of measuring the extensions of these objects such as $R_{\rm H}$, $R_{25}$ or their variants. In fact, the use of $R_{\rm 1}$ substantially modifies the scaling relations where galaxy size is an important parameter. This is particularly the case compared to  $R_{\rm e}$. We will show that the use of $R_{\rm 1}$ significantly decreases the scatter of the stellar mass-size relation by a factor of 2.5. Moreover, using  $R_{\rm 1}$, galaxies with stellar masses from 10$^{7}$ M$_\odot$ to 10$^{11}$ M$_\odot$ share the same stellar mass-size trend. The overall decrease in scatter essentially tightens the observed correlation between galaxy size and stellar mass, thus allowing us to gain insight about  the size of an object if its stellar mass is known or viceversa. We will discuss whether these findings indicate a more fundamental meaning of the new size estimator $R_{\rm 1}$ compared to the more arbitrary effective radius. We will also explore how $R_{\rm 1}$ compares  with other size indicators such as the radius enclosing half of the stellar mass ($R_{\rm e,M_{\star}}$), the Holmberg radius and $R_{\rm 23.5,i}$. 

This paper is structured as follows. In Section 2, we motivate the new size definition based on the location of the gas density threshold for star formation in galaxies. In Sections 3 and 4, we describe the data used and the selection of  targets. The methodology is described in Section 5 and our results presented in Section 6. Section 7 discusses the results obtained and they are  summarised in Section 8. Through the paper we assume a standard $\Lambda$CDM cosmology with $\Omega_m$=0.3, $\Omega_\Lambda$=0.7 and H$_0$=70 km s$^{-1}$ Mpc$^{-1}$.

\section{Towards a physically motivated definition for the size of  galaxies}

When defining a new way to measure the size of galaxies, it is important to select a physical criterion intimately linked to the way galaxies increase in extension. Galaxies are expected to grow both in stellar mass and size by two different phenomena. The first  is based on the transformation of gas into stars and the second is due to the accretion of new stars by merging and tidal interactions with other galaxies. While the merging process is stochastic and difficult to model, the transformation of gas into stars is strongly connected with the gas density of these systems. 

Above a given gas density threshold,  gas is transformed into stars. Consequently, the position of  these newborn stars is encircled by the location of such a critical gas density \citep{1968dms..book.....S,1972ApJ...176L...9Q,1980MNRAS.193..189F,1989ApJ...344..685K}. The radial location of this gas density threshold is thus suggestive of a natural way to define the size of  galaxies. This is the expectation  for the vast majority of  galaxies, i.e. those whose main channel of stellar mass growth is the transformation of gas into stars. This includes almost all the dwarf galaxies and the majority of disc galaxies where  growth by merging activity with other minor objects is  \citep[see e.g.][]{1992ApJ...389....5T}. The critical gas surface density for star formation is theoretically estimated to be $\Sigma_c\sim$3-10 $M_{\rm \odot}$ pc$^{-2}$ \citep[see e.g.][]{2004ApJ...609..667S}. If the efficiency of transforming gas into stars is not 100\%, a reasonable way of defining the size of a galaxy would be to locate a stellar mass isocontour at $\Sigma_\star \sim$1-3 $M_{\rm \odot}$ pc$^{-2}$. Such a range in surface density corresponds to an efficiency of gas-to-star transformation between $\sim$10-30\%. A way to test whether such a definition is reliable and a better proxy for the global luminous extension of galaxies compared to other size indicators such as $R_{\rm e}$ is to explore the stellar mass density at which the edges of  disc galaxies appear (i.e. the location of  their truncations). To the best our knowledge, such work has not been conducted exhaustively yet. However, we have some examples where this has been done in detail. For instance, in UGC00180, a galaxy with similar properties to M31, the truncation is located at $\sim$2.5 $M_{\rm \odot}$ pc$^{-2}$ \citep[this is an upper limit as the projection effect has not been taken into account;][]{2016ApJ...823..123T}. For another two edge-on nearby galaxies (NGC4565 and NGC5907), the stellar mass density at their truncation radii is between 1-2 $M_{\rm \odot}$ pc$^{-2}$ \citep{2019MNRAS.483..664M}. The fact that the fraction of stars beyond the truncation of NGC4565 and NGC5907 declines to  0.1-0.2\% reinforces the idea that such a stellar mass density is a good proxy for defining the luminous size of a galaxy.  This number is compatible with a tiny fraction of stars that migrated from a region within the truncation radius to the outskirts. Unlike $R_{\rm e}$, an added value to the physically motivated size definition we are proposing is that the measurement corresponds to what the human eye identifies as the border of an object.

In what follows, \textit{we propose the radial location of the gas density threshold for star formation as our size definition}. Based on theoretical arguments and observational evidence of Milky Way-like galaxies, we suggest an operative way to estimate this density threshold for star formation by using a stellar mass density isocontour at $\Sigma_\star \sim$1 $M_{\rm \odot}$ pc$^{-2}$. We  refer to the radial position of such an isomass contour as $R_{\rm 1}$. Obviously, the choice of 1 $M_{\rm \odot}$ pc$^{-2}$, instead of, for example, 0.5, 2 or 3 $M_{\rm \odot}$ pc$^{-2}$, depends on the exact efficiency of  star formation among different galaxies. Therefore, depending on the galaxies' characteristics other values could perhaps better enclose the location of  in-situ star formation. In this paper, we have preferred to adopt a relatively low efficiency in transforming gas into stars to be as  inclusive as possible. In this regard, if anything, our measure for the size of  galaxies could  lead to slightly larger sizes  where the efficiency of forming stars is higher than what we assume in this work (see Appendix \ref{starformationproxy}). On the contrary, if the star formation were very inefficient (as may well be  the case for dwarf galaxies), our measure of size will be biased towards smaller sizes \citep[see][]{chamba2020}. In Appendix  \ref{app:alternatives}, we discuss the use of  alternative proxies to locate the radial location of the gas density threshold for star formation.

 In the previous paragraphs, we have motivated the use of the radial location of the stellar mass isocontour at $\Sigma_\star \sim$1 $M_{\rm \odot}$ pc$^{-2}$ as an operative method to locate the gas density threshold for star formation and thus characterise a physical size for  galaxies. This size measure should  work particularly well for galaxies whose main growth channel is the transformation of gas into stars. What would be the plight of such a definition for spheroidal galaxies? Those galaxies are thought to form a significant fraction of their stars in an early-on intense starburst and later on add new stars (mostly to their periphery) through merging with other (satellite) galaxies. Most of this secondary growth is produced by dry minor mergers \citep[see e.g.][]{2011MNRAS.415.3903T}. As a matter of fact, we will show in this paper that the proposed size definition is useful to separate the core of spheroidal galaxies (predominantly formed by an intense star formation burst) from the material that is later on accreted by minor merging. A discussion of the limits of the new size measure is given in Appendix \ref{app:limits}.

\section{Imaging Data: the IAC Stripe82 Legacy Project}

To estimate the location of the density threshold for star formation through the position of the 1 $M_{\rm \odot}$ pc$^{-2}$ isomass contour,   a survey with multi-wavelength colour information is necessary. As we will explain in Sec. \ref{section:Method},  the stellar mass density profiles of the objects can be estimated using different combinations of optical bands. In this work, we have used the IAC Stripe82 Legacy Project (hereafter IAC Stripe82) data set \citep{2016MNRAS.456.1359F,2018RNAAS...2c.144R} as our deep imaging survey. This dataset is a co-addition of the SDSS Stripe82 data \citep{2008AJ....135..338F} with the goal of retaining the faintest surface brightness structures. The average seeing is 1 arcsec and the pixel scale is  0.396 arcsec. The total area of the survey is 275 square degrees. To conduct the present work we have used publicly available rectified images from this data  set (\url{http://research.iac.es/proyecto/stripe82/}). In addition to the imaging data, the public release also includes photometric catalogues \citep{2018yCat..74561359F}. The mean limiting surface brightness of the survey are $\mu_g$=29.1, $\mu_r$=28.6, and $\mu_i$=28.1 mag arcsec$^2$ (equivalent to a 3$\sigma$ fluctuation with respect to the background of the image in an area of 10$\times$10 arcsec$^2$).

\section{Target selection}  
\label{section:SelectionData}

Having introduced a new size definition based on the radial location of the gas density threshold for star formation, we will now explore its use across a galaxy mass range as large as possible and how it performs for different morphological types. This is relevant as the star formation history could be very different depending on the galaxy's characteristics. The cosmological volume covered by the Stripe82 data, together with its depth, thus allows us to collect a relatively large sample of galaxies  with a wide range of stellar masses and morphologies.

We have selected 1005 galaxies with z$<$0.09 spanning five orders of magnitude in stellar mass (10$^7$$M_{\rm \odot}$$<$$M_{\rm \star}$$<$10$^{12}$$M_{\rm \odot}$). This collection of galaxies extends from the dwarf galaxies regime up to giant spirals and ellipticals. All the galaxies with M$_{\star}>10^9$ $M_{\rm \odot}$ were selected from the \citet{2010ApJS..186..427N} catalogue, which includes a detailed visual classification of about 14000 galaxies in the SDSS footprint. We have selected all the galaxies listed in this catalogue that are within the Stripe82 area (i.e. 1010 objects). Unfortunately, the \citet{2010ApJS..186..427N} catalogue lacks objects with stellar masses below  10$^9$ $M_{\rm \odot}$. For this reason, to increase our sample towards 
less massive galaxies (10$^7$$M_{\rm \odot}$$<$$M_{\rm \star}$$<$3$\times$10$^{9}$$M_{\rm \odot}$), we retrieve them directly from the \citet[][]{2013MNRAS.435.2764M} catalogue. For the dwarf sample we lack morphological information. In order to have enough spatial resolution for our size analysis, we select only nearby dwarf galaxies, i.e. with 0.002$<$z$<$0.018. Within such a redshift range and  the Stripe82 area we find 323 galaxies from the \citet[][]{2013MNRAS.435.2764M} catalogue.

Of the total 1333 initially selected galaxies, 1005 were used for the final analysis, and  328 galaxies  removed for multiple reasons. In some cases, the galaxies are located very close to a bright star or galaxy (152 objects), making the retrieval of their surface brightness profile unreliable. In other cases (103 objects) the galaxies are dramatically affected by dust contamination from Galactic cirri or several neighbouring objects that crowd the outskirts of the galaxy. 48 galaxies that have an axis ratio smaller than 0.3 were also removed (See \ref{section:inclination}). 22 galaxies for which the TType was classified as `unknown' in the \citet{2010ApJS..186..427N} catalogue were discarded. And finally, 3 galaxies were removed as they appeared at the edge of the Stripe82 footprint and only part of the galaxy was visible in the images.

The  sample of massive galaxies was separated into two morphological groups depending on the TType classification by \citet{2010ApJS..186..427N}. Those galaxies with TType$>-$1 (in total  464 objects) were called ``spiral galaxies'' and contain morphologies from S0/a to Im, while those with TType$<-$1 (in total  279 objects) are dubbed ``ellipticals'' and contain the morphological classes E0 to S0+. After the cleaning process, the remaining ``dwarfs'' comprise  262 galaxies in our final sample. For completeness, in the table where we provide the properties of these galaxies, the TType for dwarf galaxies is indicated as -99 (see Sec. \ref{section:results}).

\section{Methodology}  
\label{section:Method}

As explained in the Introduction, in this paper we explore the stellar mass - size relation of  galaxies using our new size definition. In addition, we compare this mass - size relation with those resulting from the  use of traditional size measurements such as the effective radius, the half-mass radius and the Holmberg and $\mu_i$=23.5 mag/arcsec$^2$ isophotal radii. In order to estimate the structural parameters necessary for this analysis we need to conduct a number of steps that are explained in the following subsections. For all  galaxies, we create images in  the \textsl{g} and \textsl{r} filters of 600 kpc $\times$ 600 kpc in size in the rest-frame of each galaxy and centred on the object of interest. The pipeline developed for this work is written using  \texttt{Python v. 3.6.5}\footnote{\protect \url{https://www.python.org/}}.

\subsection{Removal of scattered light from point sources and masking}

The scattered light from bright stars was modelled and subtracted using the procedure of \citet{2016ApJ...823..123T}. This is a key step that is necessary to explore  low surface brightness features with confidence \citep[see e.g.][]{1991ApJ...369...46U,2009PASP..121.1267S}. All  stars brighter than 17 mag were identified using the $G$-band reported in the GAIA DR 1 catalogue \citep{2016gaia}. To produce the scattered light field, we use the extended (radial size of $\sim$8 arcmin) point spread function (PSF) models in all the SDSS bands created by \citet{2020MNRAS.491.5317I}. The pipeline to remove the scattered light from point sources in the IAC Stripe82 fields will be fully described in a future publication and applied to the full IAC Stripe82 survey (N. Chamba et al. in prep.).

Once the scattered light is removed from the images, it is necessary to mask all  remaining sources that are affecting the light distribution of the galaxy we are exploring. To conduct this task, we used a Python implementation of \texttt{MTOBjects} \citep{2016mto}, a tree-based detection scheme which is robust against false positives, especially important for the identification of extended low signal-to-noise structures in deep imaging (C. Haigh et al. submitted). For this work, the algorithm parameter \texttt{move\_up = 0.3} and the $\alpha$ parameter for statistical testing was set to its default value. 

\subsection{The effect of inclination}
\label{section:inclination}

The surface brightness of  galaxies (particularly those following a disc-like configuration) are strongly affected by the inclination of the object. The larger the inclination of a galaxy, the brighter it appears to an observer as the number of stars along the line of sight increases. As our proposed size definition requires a proper estimation of the flux in the outer regions of galaxies, we correct the brightness of  galaxies by the effect of its inclination. This is  not straightforward  and has been investigated in-depth in multiple papers \citep[see e.g.][]{1958MeLuS.136....1H,1975gaun.book..123H,1985ApJS...58...67T,1994AJ....107.2036G}. To estimate the correction we need to apply to the data, we  build a 3D disc model assuming an exponential decline for the radial light distribution \citep{1959HDP....53..275D,1970ApJ...160..811F} and a sech$^2$ in the vertical $z$ direction. This is expected for an isothermal population in a plane-parallel system \citep{1942ApJ....95..329S,1950MNRAS.110..305C}. The luminosity distribution of the model is:

\begin{equation}
L(R,z)=L_0\exp\bigg(\frac{-R}{h}\bigg)sech^2\bigg(\frac{z}{2z_0}\bigg)
\end{equation}

where $L_0$ is the central luminosity density, $h$ is the scale length and $z_0$ is the scale height. The model was created using \texttt{IMFIT} \citep[we used the model \texttt{ExponentialDisk3D} with \texttt{n=1};][]{2015ApJ...799..226E}. We probe three different models with $z_0/h$=0.08, 0.12 and 0.17. The ratio of these parameters covers the values measured for the thin disc of our own Milky Way and its uncertainties \citep{2016ARA&A..54..529B}. This model is an idealised version of discs. Real discs are much more complex, containing clumps, dust, warps, etc. In addition, we have not considered possible corrections due to internal dust. Therefore, any dependence of the model on  wavelength is neglected \citep[for a detailed analysis of this issue see][]{2019arXiv190901572K}.

\begin{figure}
\includegraphics[width=90mm]{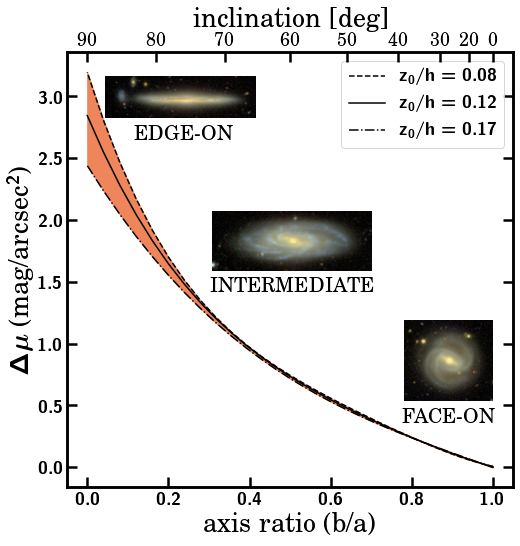}

\caption{Difference in surface brightness $\Delta \mu=\mu_{\rm face-on}-\mu_{\rm inc}$ between the face-on orientation and a given inclination for a disc-like galaxy. The figure shows $\Delta \mu$ along the semi-major axis of a galaxy model at a radial distance $R=5h$ and is explored for three different disc thicknesses (shown in the legend) parameterized using $z_0/h$.}
\label{fig:inclination}
\end{figure}

Since we are mostly interested in the effect of  inclination on the brightness of the intermediate-outer regions of  galaxies, we calculate the difference in surface brightness ($\Delta\mu$) at a given inclination $i$ ($\mu_{\rm inc}$) compared to the face-on orientation ($\mu_{\rm face-on}$) at a radial distance of $R=5h$ (i.e. $\Delta\mu$=$\mu_{\rm face-on}$-$\mu_{\rm inc}$). The difference $\Delta\mu$ was estimated at all inclinations along the semi-major axis of the model galaxy (see Fig. \ref{fig:inclination}).

Figure \ref{fig:inclination} shows that the difference in brightness $\Delta\mu$ produced by different disc thicknesses (as parameterized by $z_0/h$) is only noticeable at very large inclinations (i.e. $i>$70 degrees). For this reason, we remove any galaxy with an axis ratio smaller than 0.3 from the sample (Sec \ref{section:SelectionData}). To facilitate the reader with the application of this inclination correction, we provide in Table \ref{table:deltamu} the values of the coefficients of a polynomial fit to the different models shown in Fig. \ref{fig:inclination}:

\begin{equation}
\Delta\mu=\sum_{j=0}^{4}\alpha_j(b/a)^j
\end{equation}

with $b/a$=cos($i$), the ratio of the semi-minor to the semi-major axis of the isophote used to measured the inclination of  galaxies. The polynomial fit we provide is very accurate, with an error in $\Delta\mu<0.01$ mag. In the next Section, we explain how to apply this correction to real data. In this work, we have used the correction corresponding to the ratio $z_0/h$=0.12.

\begin{table*}
\centering
\caption{Values of the polynomial coefficients to correct the surface brightness profiles of disc-like galaxies for the inclination effect. The coefficients have been calculated for three different disc thicknesses which are parameterised by the ratio  $z_0/h$. As we are interested in the effect of inclination on the intermediate and outer regions of  galaxies, this correction has been estimated at a distance $R=5h$ (see text for details).}

\label{table:deltamu} 
\begin{tabular}{|c|c|c|c|c|c|}
\hline 
\textbf{$z_0/h$} & \textbf{$\alpha_0$} &\textbf{$\alpha_1$}&\textbf{$\alpha_2$} &\textbf{$\alpha_3$} & \textbf{$\alpha_4$}  \\  

\hline 

 0.08 & 3.195 &  -10.396 &   17.584  &   -16.033 &  5.657 \\
 0.12 & 2.845 &  -7.833 & 10.792 &  -8.482 &   2.679  \\
 0.17 & 2.440 &  -5.273 &  4.577 &   -1.932 & 0.185 \\

\hline 
\end{tabular}
\end{table*}

\subsection{Stellar mass density profiles}

After the removal of  scattered light and masking the images, we extract the surface brightness profiles of the galaxies in the \textsl{g} and \textsl{r} bands to obtain their stellar mass density profiles. The surface brightness profiles are obtained using elliptical apertures with a fixed centre, axis ratio $q$ and position angle (PA). As a first guess for the centre of the galaxies, we use the R.A. and Dec information provided by the SDSS catalogues.

To determine the axis ratio and PA, for each galaxy we use those pixels where the surface brightness is between 25 and 26 mag/arcsec$^2$ in the $g$-band. The spatial distribution of these pixels were fit to an ellipse. The PA (in degrees) is the angle between the semi-major axis and the horizontal axis, measured in the counter clockwise direction from the horizontal axis.  The fit parameters (centre, axis ratio and PA) of the ellipses were then visually checked to ensure the outermost parts of the galaxies were characterised properly. If not, they are corrected accordingly. Once fixed, surface brightness profiles of the galaxies are extracted by averaging their flux over annuli parameterised by the fit ellipse. These profiles are extracted up to a radial distance of 200 arcsec which is well beyond the visual extension of our galaxies. This is crucial in order to retrieve a sensible characterisation of the outer part of  galaxies, particularly when the criterion we are proposing in this work is based on the location of a low stellar mass density contour such as 1  $M_{\rm \odot}$ pc$^{-2}$. 

Another important effect that must be accounted for when obtaining surface brightness profiles is defining the level of the background. Although the IAC Stripe82 images we have used are background subtracted, in some occasions the subtraction was not precise enough to be a reliable representation of the (local)  surrounding background value of the galaxies (i.e. a slight under- or overestimation). For this reason, in order to have the most accurate background subtraction as possible, we followed the procedure developed by \citet{2006A&A...454..759P}. The radial distance up to which the profiles have been extracted (i.e. 200 arcsec) is about two times the location of the isophote at 26.5 mag/arcsec$^2$ (\textsl{r}-band) in the case of ellipticals and three times for the  spiral and dwarf galaxies. This allows us to determine the background brightness in regions very close to the galaxies by identifying the asymptotic value in the number of counts around the object. We fit that value, subtract/add it to the images and obtain the profile once more. 

We then correct the surface brightness profiles for Galactic extinction. The extinction corrections A$_g$ and A$_r$ are obtained from NED taking into account the location of each galaxy on the sky (\url{https://ned.ipac.caltech.edu/forms/calculator.html}). Following this, the effect of the inclination (see Section \ref{section:inclination}) is corrected for spiral and dwarf galaxies  as follows. For each galaxy we measure its inclination based on the axis ratio we have determined before. The inclination correction, $\Delta\mu$, is then directly applied to the derived surface brightness profiles. The same inclination correction is applied for both \textsl{g} and \textsl{r} profiles, therefore the colour radial profile of galaxies remains unaffected. Due to our limited photometric information, we do not attempt any correction for internal dust. 

The final step is to obtain the stellar mass-to-light ratio ($M/L$) profile.  Once the $M/L$ is known, the following equation \citep[see e.g.][]{2008ApJ...683L.103B}:

\begin{equation}
\label{equ1}
\log\Sigma_\star = \log (M/L)_\lambda - 0.4(\mu_\lambda-\mu_{\rm abs,\odot,\lambda})+8.629
\end{equation}

where $\mu_{\rm abs,\odot,\lambda}$ is the absolute magnitude of the Sun at wavelength $\lambda$, is used to obtain the stellar mass density (in $M_{\rm \odot}$ pc$^{-2}$) as a function of the surface brightness.

 To compute  $M/L$, we followed the procedure described by \citet{2015MNRAS.452.3209R}. As a basis for our estimation, we used the $g-r$ colour and the surface brightness in the \textsl{g}-band. We use the parameters provided by \citet{2015MNRAS.452.3209R} that correspond to the \citet[][BC03]{2003MNRAS.344.1000B}  models and a Chabrier IMF \citep{2003PASP..115..763C}. 
 
 Despite the obvious advantage of decreasing the effect of galactic dust by using the \textsl{i}-band instead of the bluer \textsl{g} and \textsl{r} bands, we prefer to use the latter filters to estimate our size indicator for two main reasons. Firstly, the sky brightness in the \textsl{i}-band is around two magnitudes brighter than in the \textsl{g}-band \citep[see e.g. Fig. 1 in ][]{2012MNRAS.425.2741H}. This effect is not compensated by the brighter emission of the stellar populations towards the red (which is typically between \textsl{g-i}=0.5 to 1 mag for spiral galaxies). As  a result,
 the  redder SDSS bands are noisier at a given surface brightness because all the  SDSS bands have the same  integration time. Secondly,  as our size indicator is estimated through a colour combination,  the effect of the
PSF on the surface brightness profiles should not be very different from band to band. This applies for
\textsl{g} and \textsl{r}, but in the case of the \textsl{i}-band, the SDSS PSF is significantly different for those in \textsl{g} and \textsl{r}, as  can be seen in 
\citet[][Figure 2]{2008MNRAS.388.1521D} and \citet[][Figure 8]{2020MNRAS.491.5317I}.

\subsection{Estimating the structural parameters of galaxies}
\label{sec:parameters}

Once the stellar mass density profiles of the galaxies are created, it is straightforward to obtain the total stellar mass and the location of $R_{\rm 1}$, the proxy for the location of the gas density threshold for star formation we have adopted as a measure of size in this work. This procedure has been performed  for all the galaxies in our sample. In order to get a homogeneous determination of the total stellar mass of all our galaxies, we have integrated their mass density profiles. The integration takes into account the axis ratio of the galaxy and therefore assumes an elliptical symmetry for the distribution of  light, from the central position of the object up to the radial location provided by the 29 mag/arcsec$^2$ isophote (\textsl{g}-band). This estimate of the total stellar mass is a lower limit to the total mass  of the object. However, the limiting isophote we are using is extremely faint, therefore the amount of stellar mass beyond such an isophote is expected to be very low \citep[$<$3\%;][]{2001MNRAS.326..869T}. We prefer to use this approach for estimating the total stellar mass instead of assuming a shape for the light distribution (i.e. exponential, de Vaucouleurs, etc) and extrapolating the stellar mass density profiles to infinity. In  Appendix \ref{app:masscomparison}, we compare our stellar mass determination with that of the Portsmouth Spectro-Photometric Stellar Mass computation \citep[][]{2013MNRAS.435.2764M} and find that both mass determinations are  similar. Finally, we determine the location of $R_{\rm 1}$  directly using the stellar mass density profiles. Estimates of the half-mass radii are done using the cumulative mass density profiles.

The  effective radii of  galaxies are determined from the \textsl{g}-band  images (our deepest data)\footnote{To check the robustness of our estimation of $R_{\rm e}$ using the \textsl{g}-band, we also estimated the same quantity using the \textsl{i}-band. We found a very tight correlation between both effective radii (Pearson correlation coefficient r=0.996). As expected, we find that R$_{e,g}$ is slightly
larger than R$_{e,i}$: Re,g/Re,i=1.030$\pm$0.002, with a dispersion of 0.083. Both effective
radii are thus very similar.}. We use the growth curve in  \textsl{g} to obtain the radial location within which half of the total light of the galaxy is contained. As we have done for the total stellar mass,  the total light of the galaxy is measured as the light enclosed by the observed 29 mag/arcsec$^2$ isophote (\textsl{g}-band). By definition, $R_{\rm e}$ is not affected by  Galactic extinction  nor the inclination correction of the profiles except indirectly for the location of the 29 mag/arcsec$^2$ isophote (\textsl{g}-band). In addition, we also estimate  the Holmberg Radius ($R_{\rm H}$) for all our galaxies. Lacking the \textsl{B}-band in our survey,  the location of the observed isophote at $\mu_g$=26 mag/arcsec$^2$ was considered as a proxy for $R_{\rm H}$. Using the \textsl{i}-band profiles, we also determine the radial location of the isophote corresponding to 23.5 mag/arcsec$^2$ (i.e. $R_{\rm 23.5,i}$). Both isophotal sizes were estimated after correcting the profiles for Galactic extinction and cosmological dimming. All these structural parameters of our galaxies are provided in Table \ref{table:parameters}.

\begin{landscape}
\begin{table}
\centering
\caption{Characteristics of the galaxies used in this work. Unless explicitly stated otherwise, the quantities provided in this table have been derived in this work. This table includes the name of the galaxies \citep{2018ApJS..235...42A}, their spatial location \citep{2018ApJS..235...42A}, their axis ratio ($q$) and the position angle of the ellipses (PA) used to extract the surface brightness profiles (measured counter-clockwise starting from the horizontal axis),  Galactic extinctions  in the \textsl{g},  \textsl{r} and \textsl{i} bands (from NED),  spectroscopic redshift ($z$) \citep{2018ApJS..235...42A},  morphological type \citep[][TType=-99 corresponds to dwarf galaxies]{2010ApJS..186..427N},  effective radius $R_{\rm e}$ (measured in the \textsl{g}-band), the half-mass radius ($R_{\rm e,M_{\star}}$), the radius corresponding to the location of the isophote with $\mu_i$=23.5 mag/arcsec$^2$ (i.e. $R_{\rm 23.5,i}$), the Holmberg Radius ($R_{\rm H}$), the radial location  $R_{\rm 1}$ of the isomass contour at 1 $M_{\rm \odot}$ pc $^{-2}$ and the stellar mass of  galaxies (assuming a Chabrier IMF). The quantities are given showing only the significant figures up to which the values can be regarded reliable. The table shows those galaxies in Fig. \ref{fig:examples} and \ref{fig:insituexsitu}, in order of appearance. The complete table is available in the online version of the paper.}

\label{table:parameters} 
\begin{tabular}{|c|c|c|c|c|c|c|c|c|c|c|c|c|c|c|c|}
\hline 
\textbf{JID} & \textbf{R.A.} & \textbf{Dec}&\textbf{$q$} &\textbf{PA} & \textbf{A$_g$}  & \textbf{A$_r$} & \textbf{A$_i$} & \textbf{$z$} & \textbf{TType} & \textbf{$R_{\rm e}$} & \textbf{$R_{\rm e,M_{\star}}$} & \textbf{$R_{\rm 23.5,i}$} & \textbf{$R_{\rm H}$} & \textbf{$R_{\rm 1}$} & \textbf{Log(M$_\star$/$M_{\rm \odot}$)}  \\  
 & \textbf{(deg)} &\textbf{(deg)}& &\textbf{(deg)} & \textbf{(mag)}  & \textbf{(mag)} & \textbf{(mag)} &  &  & \textbf{(kpc)} & \textbf{(kpc)} & \textbf{(kpc)} &  \textbf{(kpc)} & \textbf{(kpc)} & \\  

\hline

J010301.72-010639.46 & 15.75723 & -1.11113 & 0.88 & 95.0 & 0.134 &  0.093 & 0.069 & 0.0175 &  -2 & 2.45 & 3.67 & 7.94 & 11.67 & 18.56 & 10.38 \\ 
J005753.69-004852.90 & 14.47382 & -0.81479 & 0.96 & 120.0 & 0.094 & 0.065 & 0.048  & 0.0419 & -2 & 2.78  & 4.47 & 12.41 & 17.27 & 25.08 & 11.03 \\
J000150.32+010155.24 & 0.45973 & 1.03172 &  0.49 & 179.0 & 0.085 & 0.059 & 0.044  & 0.0862 & -5 & 24.99  & 29.13 & 46.05 & 71.98 & 148.12 & 11.82 \\
J003934.82+005135.83 & 9.89529 & 0.85979 &   0.71 & 61.0  &  0.066 & 0.046 &  0.034 & 0.0146 & 5  & 8.64  & 7.70 & 16.39 & 20.87 & 23.01 & 10.37 \\ 
J012223.77-005230.73 & 20.59913 & -0.87523 &  0.44 & 42.0  & 0.169 & 0.117 & 0.087 & 0.0271 & 4 &  16.17 & 9.65 & 32.27 & 44.35 & 44.04 & 10.96 \\ 
J021219.69-004841.46 & 33.08210 &  -0.81153 &  0.89 & 37.0 & 0.097 & 0.067 & 0.050 &  0.0408 &  0 &  9.18 & 5.90 & 23.68 & 32.61 & 37.17 & 11.24 \\ 
J021808.12+004529.8  & 34.53385 &  0.75830  & 0.64 & 133.0 & 0.130 & 0.090 &  0.067 & 0.0092 &  -99 & 1.00 & 1.49 & 1.79 & 2.86 & 2.97 & 8.05 \\   
J233646.86+003724.2  & 354.19526 &  0.62341 &   0.86 & 88.0 &  0.113 & 0.078 & 0.058 &  0.0088 & -99 & 2.13 & 2.38 & 2.58 &  4.44 &  5.44 &  8.65 \\ 
J010607.19+004633.5  & 16.52997 &  0.77599 &   0.84 & 48.0 &  0.084 & 0.058 & 0.043 & 0.0174 &  -99 & 5.43 & 5.21 & 4.14 & 10.87 & 10.91 & 9.10 \\
J235618.80-001820.17 & 359.07860 & -0.30583 & 0.71 & 99.0 & 0.124 & 0.086 & 0.064  & 0.0241 & -3 & 4.11 & 5.08 & 16.70 & 26.56 & 48.39 & 11.15 \\ 
J024331.30+001824.49 & 40.88040 &  0.30676 &   0.36 & 108.0 & 0.124 & 0.086 &  0.064 &  0.0267 & 3 &  8.82 & 5.72 & 22.89 & 30.75 & 29.43 & 10.57 \\ 

\hline 
\end{tabular}
\end{table}
\end{landscape}

\section{Results} \label{section:results}

\begin{figure*}
\includegraphics[width=\linewidth]{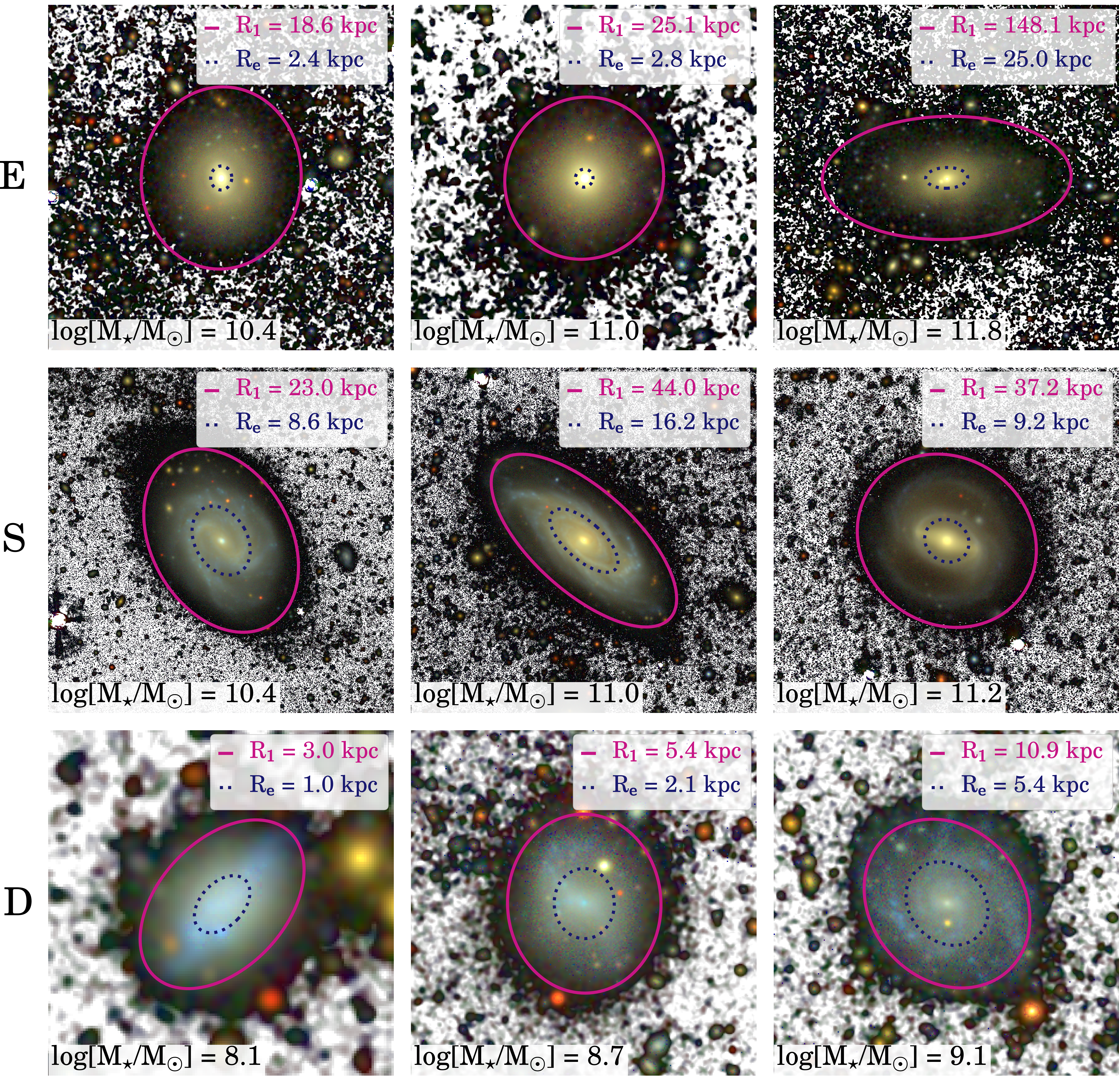}

 \caption{Collection of galaxy images showing the location of their  $R_{\rm e}$ and $R_{\rm 1}$, the isomass contour at 1 $M_{\rm \odot}$ pc$^{-2}$.  The top row shows galaxies which have been classified as ellipticals (E0-S0+), the middle row shows spiral (S0/a-Sm) galaxies, and the lower row shows dwarf galaxies. The galaxies are displayed with increasing stellar mass from left to right. This figure clearly illustrates how the proxy for the gas density threshold for star formation ($R_{\rm 1}$) nicely encloses the bulk of the stellar mass of  galaxies. The coloured regions of the images are the IAC Stripe82 \textsl{g, r} and \textsl{i} band composite, while the black and white background is the sum of the 3 bands. The background level of these images is $\sim$29.1 mag/arcsec$^2$ (3$\sigma$ 10$\times$10 arcsec$^2$; \textsl{r}-band).  Table \ref{table:parameters} lists details of the galaxies shown.}
\label{fig:examples}
\end{figure*}

\begin{figure*}
	\centering
    \includegraphics[width=\linewidth]{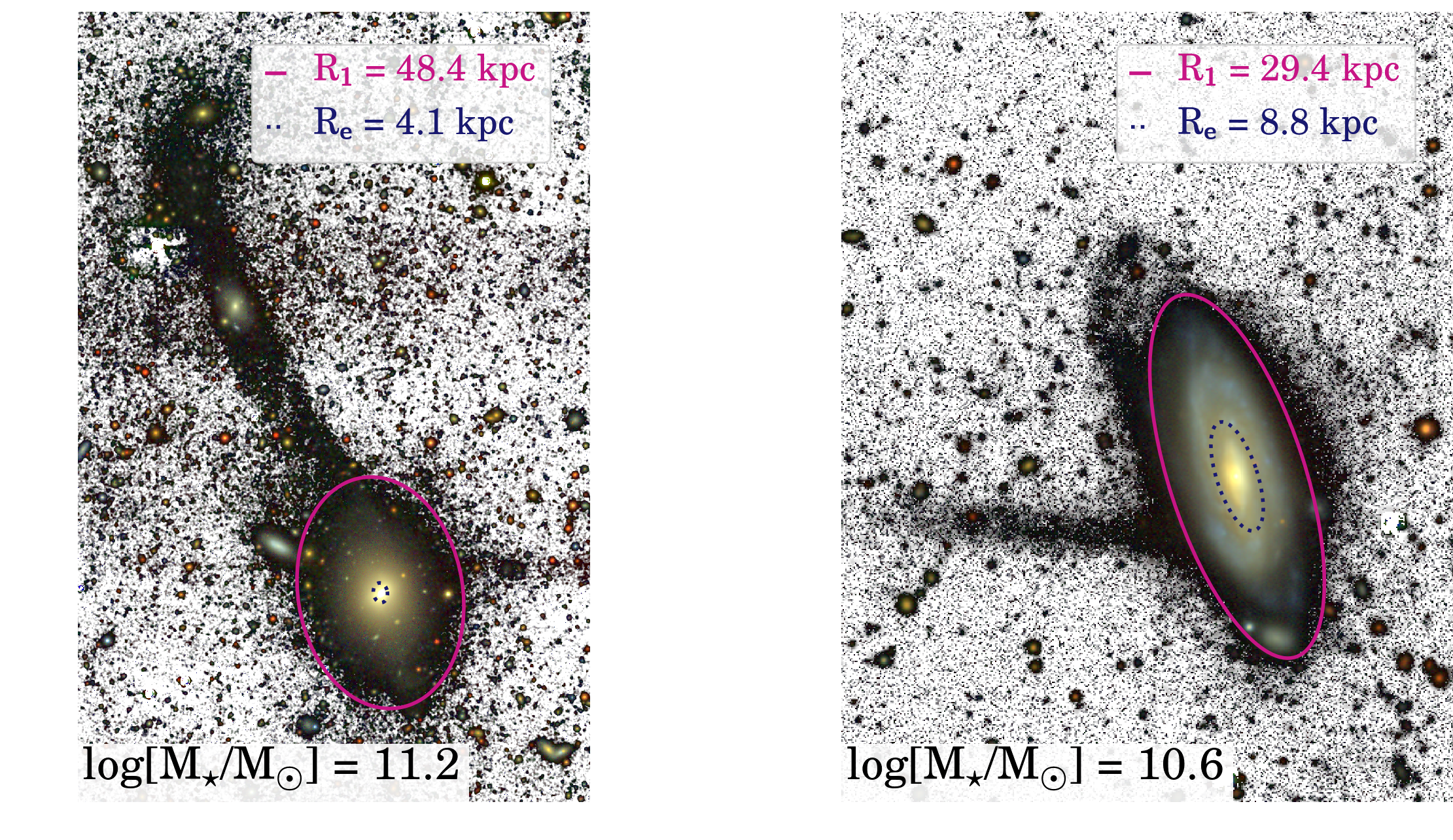}

\caption{Separating \textit{in-situ} star formation from \textit{ex-situ} stellar accretion using our size measure (see N. Chamba et al. in prep. for further details). The two images show how the isomass contour at 1 $M_{\rm \odot}$ pc$^{-2}$ (i.e. $R_{\rm 1}$ or the proxy for the location of the gas density threshold for star formation) nicely divides the structure of  galaxies into two different parts: the inner region where the bulk of the stars is contained and where  \textit{in-situ} star formation is or has been taking place and the external region where streams of on-going accretion are clearly visible. The left panel shows a massive elliptical and the right panel corresponds to a spiral galaxy with similar stellar mass to the Milky Way. See Table \ref{table:parameters} for details on the galaxies shown.}
\label{fig:insituexsitu}
\end{figure*}

Figure \ref{fig:examples} shows a few representative galaxies in our sample to illustrate the difference between the location of  their $R_{\rm 1}$ and $R_{\rm e}$ contours. The size based on the location of the gas density threshold for star formation much better represents the intuitive concept of the size of  galaxies, such as its edge or boundary compared to $R_{\rm e}$. Expanding on this point, Fig. \ref{fig:insituexsitu} shows the location of $R_{\rm 1}$ and $R_{\rm e}$ for two galaxies with clear signatures of on-going stellar accretion. In these examples, the location of $R_{\rm 1}$ may serve as a marker to separate the stellar material which is in the form of streams (formed \textit{ex-situ}) from those stars which are located in the bulk (\textit{in-situ}) of the main galaxy. An in-depth analysis of the use of $R_{\rm 1}$ (and its variants) for this purpose will be presented in a future publication (N. Chamba et al. in prep).

\subsection{The properties of the $R_{\rm 1}$ - mass relation}

\begin{figure*}
\includegraphics[width=0.9\linewidth]{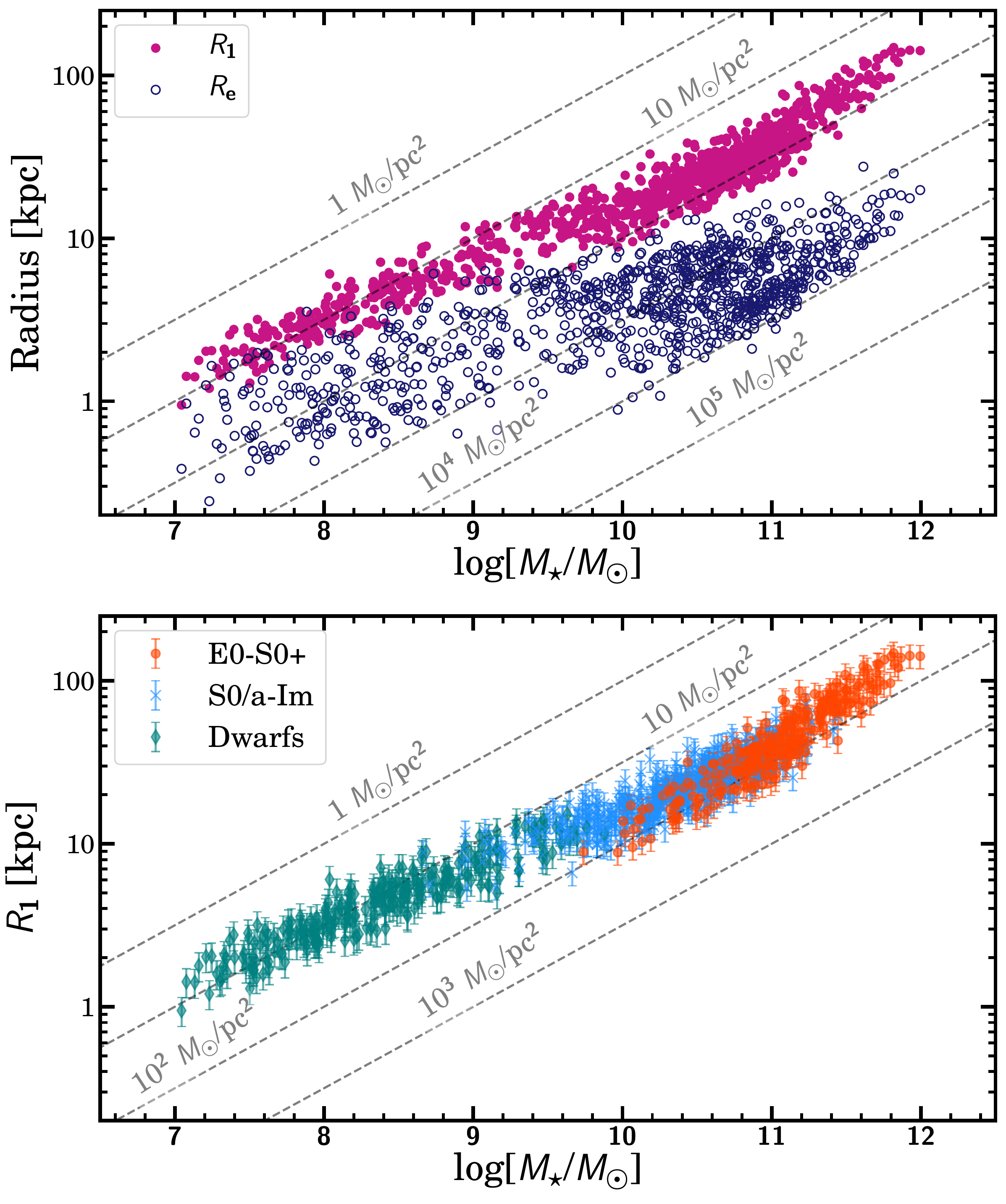}
\caption{Stellar mass - size relation for the galaxies in our sample. Upper panel: The observed $R_{\rm 1}$ - mass  and  $R_{\rm e}$ - mass relations, where $R_{\rm e}$ has been measured using the \textsl{g}-band. The scatter of the relation using $R_{\rm 1}$ is significantly smaller compared to that with $R_{\rm e}$. Lower panel: The same $R_{\rm 1}$ - mass relation after splitting our sample into three categories: ellipticals (E0-S0+), spirals (S0/a-Im) and dwarfs as labelled in the legend. Spiral and dwarf galaxies follow the same trend, while massive ellipticals with $M_\star$>10$^{11}$ $M_{\rm \odot}$ show a tilt with respect to less massive galaxies. The grey dashed lines correspond to locations in the plane with constant (projected) stellar mass density.}
\label{fig:scale_relation}
\end{figure*}

The main result of this paper is shown in Fig. \ref{fig:scale_relation}: the mass - size relation spanning over five orders of magnitude in stellar mass (10$^7$-10$^{12}$ $M_{\rm \odot}$). The figure illustrates how the mass - size relation changes when using $R_{\rm 1}$ instead of $R_{\rm e}$ as a size measurement of  galaxies. To extract both the slope and dispersion of the relations, we used a Huber Regressor \citep{huber1964}, which is a linear regression model that is robust to outliers. We list a number of enlightening results:

\begin{itemize}
\item $R_{\rm 1}$ is a factor of 5 to 10 larger than $R_{\rm e}$ in all  galaxies.
\item  The observed scatter of the stellar mass - size relation is significantly lower  by a factor of $\sim$2 (from $\sigma_{\rm R_e}$$\sim$0.17 dex to $\sigma_{R_1}$$\sim$0.09 dex)
compared to the scatter using  $R_{\rm e}$ as a size indicator. As we will show in the next section, once the observational and methodological uncertainties are accounted for, the scatter of the stellar mass - size relation drops even more to a tiny 0.06 dex (i.e.  a factor of $\sim$2.5 smaller than the intrinsic scatter using $R_{\rm e}$). The observed global scatter of the R1 - mass size relation is also lower than the observed one found using other popular size estimators ($\sigma_{R_{e,M_{\star}}}$$\sim$0.12 dex, $\sigma_{R_H}$$\sim$0.11 dex and $\sigma_{R_{23.5,i}}$$\sim$0.11 dex; see Table \ref{table:fitting}).

\item The average 2D stellar density (as measured within $R_{\rm 1}$) changes from $\sim$10 $M_{\rm \odot}$ pc$^{-2}$  for the less massive galaxies  to  $\sim$100 $M_{\rm \odot}$ pc$^{-2}$ for the most massive spiral galaxies. Above 10$^{11}$ $M_{\rm \odot}$, the average 2D stellar density of the galaxies decreases again.
\item From  10$^7$ to 10$^{11}$ $M_{\rm \odot}$ all  galaxies are located on the same mass - size relation following a power law, $R_{\rm 1}$$\propto$M$_\star$$^{\beta}$, with $\beta$=0.35$\pm$0.01. This value is compatible with the one found by \citet{2012MNRAS.425.2741H} who compared the disc scale lengths of spiral galaxies with their luminosities and found $\beta$=0.377$\pm$0.007. Interestingly, $\beta\sim$1/3  would correspond to almost the same 3D stellar mass density ($\sim$4.5$\times$10$^{-3}$ $M_{\rm \odot}$ pc$^{-3}$) if all the stars were distributed in a sphere of radius $R_{\rm 1}$.
\item Above 10$^{11}$ $M_{\rm \odot}$, the slope of the relation rises  to $\beta$=0.58$\pm$0.02. This likely indicates that the most massive galaxies have formed or gained their stars very differently compared to galaxies with lower masses. 
\end{itemize}

In Table \ref{table:fitting} we show the best-fit parameters to a power law of the form R$\propto$M$_{\star}^{\beta}$ using  all the galaxies in the size - stellar mass relation as well as separate fits using each subsample (i.e. dwarfs, S0/a-Sm and E0-S0+). We have performed our  analysis using the new size indicator $R_{\rm 1}$ as well as other popular size indicators: $R_{\rm e}$, $R_{\rm e,M_{\star}}$, $R_{\rm H}$ and  $R_{\rm 23.5,i}$. The uncertainties in the best fit slope $(\beta)$ and dispersion of the observed relations ($\sigma_{\rm R_{obs}}$) are computed using a simple bootstrap method. One third of the measured points on the relation were randomly selected and fit at each iteration, for 1000 iterations. The spread in the distribution of the fits from this exercise is what is reported as the uncertainty in $\beta$ and $\sigma$. As we mentioned above, the observed global scatter of the $R_{\rm 1}$ - mass relation is significantly smaller than the one observed using $R_{\rm e}$ and lower than the observed scatter with all other size estimators, i.e. $R_{\rm e,M_{\star}}$, $R_{\rm H}$ and  $R_{\rm 23.5,i}$. The values of the scatter in the  size - mass relations are, however, affected by uncertainties in estimating the stellar mass and the background around the galaxies. In order to quantify how these uncertainties affect the observed scatter of the different relations and therefore compare the intrinsic scatter of the relation using $R_{\rm 1}$ with the other size indicators, we have conducted a number of tests which we describe in the next subsection.

\begin{table*}
\centering
\caption{Best fit power law slope $\beta$ and the observed dispersion of different size - mass relations. We provide the values for the entire sample as well as for the different families of galaxies. The third column corresponds to the Pearson r correlation coefficient. We also calculate the contribution to the observed scatter produced by the uncertainty in the background of the images ($\sigma_{\rm R_{back}}$) and in our stellar mass estimation ($\sigma_{\rm R_{mass}}$). The last column shows the intrinsic scatter of the size - mass relation ($\sigma_{\rm R_{int}}$) after accounting for the uncertainty in  the background and the stellar mass of the objects.}
\label{table:fitting} 
\begin{tabular}{|c|c|c|c|c|c|c|}
\hline 
Galaxy Type & $\beta$ & $\sigma_{\rm R_{obs}}$ &  r &  $\sigma_{\rm R_{back}}$ &  $\sigma_{\rm R_{mass}}$ & $\sigma_{\rm R_{int}}$  \\  
 
\hline

& & & $R_{\rm 1}$-stellar mass & & & \\

\hline

All & 0.365$\pm$0.005 & 0.089$\pm$0.005 & 0.971 & 0.045$\pm$0.003 & 0.047$\pm$0.003 &  0.061$\pm$0.005  \\
E0-S0+ & 0.580$\pm$0.022 & 0.090$\pm$0.006 &  0.936 & 0.060$\pm$0.004 &0.054$\pm$0.004 & 0.040$\pm$0.006  \\
S0/a-Sm & 0.332$\pm$0.014 & 0.089$\pm$0.005 &  0.881 & 0.045$\pm$0.003 & 0.035$\pm$0.002 & 0.068$\pm$0.005 \\
Dwarfs & 0.362$\pm$0.016 & 0.088$\pm$0.006 & 0.931  & 0.020$\pm$0.003 & 0.056$\pm$0.006 &  0.065$\pm$0.006   \\

\hline

& & & $R_{\rm e}$-stellar mass & & & \\

\hline

All &  0.247$\pm$0.011 & 0.168$\pm$0.009 & 0.811 & $\sim$0.001 & 0.067$\pm$0.005  &   0.154$\pm$0.009   \\
E0-S0+ & 0.553$\pm$0.032 & 0.108$\pm$0.009 & 0.894 & $\sim$0.001 & 0.092$\pm$0.006 &  0.057$\pm$0.009    \\
S0/a-Sm &  0.225$\pm$0.026 & 0.162$\pm$0.009 & 0.556 & $\sim$0.001 & 0.047$\pm$0.002 & 0.155$\pm$0.009   \\
Dwarfs & 0.283$\pm$0.040 & 0.221$\pm$0.012 & 0.621 & $\sim$0.001 & 0.064$\pm$0.004 & 0.212$\pm$0.012\\

\hline

& & & $R_{\rm e,M_{\star}}$-stellar mass & & & \\

\hline

All &  0.204$\pm$0.006 & 0.117$\pm$0.008 & 0.854 & $\sim$0.001 & 0.052$\pm$0.003  &  0.105$\pm$0.008  \\
E0-S0+ & 0.509$\pm$0.021 & 0.086$\pm$0.007  & 0.918 & $\sim$0.001 & 0.075$\pm$0.006 &  0.042$\pm$0.007 \\
S0/a-Sm &  0.196$\pm$0.022 & 0.120$\pm$0.009 & 0.595 & $\sim$0.001 & 0.040$\pm$0.002 & 0.113$\pm$0.009  \\
Dwarfs & 0.175$\pm$0.027 & 0.139$\pm$0.008  & 0.638 & $\sim$0.001  & 0.038$\pm$0.003 & 0.134$\pm$0.008\\

\hline

& & & $R_{\rm H}$-stellar mass & & & \\
\hline

All &  0.304$\pm$0.007 & 0.109$\pm$0.005  & 0.940 & 0..003$\pm$0.001 & 0.056$\pm$0.004  &  0.094$\pm$0.005 \\
E0-S0+ & 0.478$\pm$0.017 & 0.065$\pm$0.005  & 0.950 & 0.004$\pm$0.001 & 0.026$\pm$0.003 &  0.059$\pm$0.005  \\
S0/a-Sm &  0.266$\pm$0.014 & 0.105$\pm$0.005 & 0.783  & 0.003$\pm$0.001 & 0.035$\pm$0.004 & 0.099$\pm$0.005 \\
Dwarfs & 0.307$\pm$0.028 & 0.148$\pm$0.008 & 0.790 & 0.002$\pm$0.001 & 0.095$\pm$0.014 & 0.113$\pm$0.008 \\

\hline

& & & $R_{\rm 23.5,i}$-stellar mass & & & \\
\hline

All &  0.330$\pm$0.006 & 0.106$\pm$0.006 & 0.944 & 0.002$\pm$0.001 & 0.062$\pm$0.005  &  0.086$\pm$0.005  \\
E0-S0+ & 0.443$\pm$0.013 & 0.056$\pm$0.005 & 0.956 & 0.002$\pm$0.001 & 0.021$\pm$0.002 &  0.052$\pm$0.005  \\
S0/a-Sm &  0.285$\pm$0.016 & 0.104$\pm$0.006 & 0.804 & 0.003$\pm$0.001 & 0.052$\pm$0.005 & 0.090$\pm$0.006  \\
Dwarfs & 0.354$\pm$0.026 & 0.143$\pm$0.010 & 0.836 & 0.001$\pm$0.001 &0.097$\pm$0.012 & 0.105$\pm$0.010 \\

\hline 
\end{tabular}
\end{table*}


\subsection{The intrinsic scatter of the $R_{\rm 1}$ - mass  relation}
\label{sec:intrinsic}

There are two main sources of uncertainty which affect the observed scatter in the size - mass relations. The first  is the accuracy in measuring the background level around the galaxies. For some  galaxies (particularly the massive ellipticals or those with red stellar populations) the surface brightness at which $R_{\rm 1}$ is measured is very faint ($\mu_g\sim$28 mag/arcsec$^2$) and therefore, a slight under- or over- subtraction of the background would bend  the surface brightness profiles of these objects and move the location of $R_{\rm 1}$. To quantify how  this can affect the position of $R_{\rm 1}$ and the rest of the size indicators, we have taken all our observed surface brightness profiles and randomly subtracted/added a number of counts compatible with the uncertainty in  the background level around each galaxy. This variation of the background allows us to measure the variation in size for each galaxy which can then be used to estimate its contribution to the total observed dispersion in the size-mass plane. This contribution  ($\sigma_{\rm R_{back}}$) is shown in Table \ref{table:fitting}. The background determination affects  the size determination for massive ellipticals more than for spirals and/or small dwarfs. This is because the latter are mainly star-forming objects and therefore the surface brightness at which $R_{\rm 1}$ is located is brighter ($\mu_g$$\sim$26-27 mag/arcsec$^2$). This explanation also applies for the isophotal sizes $R_{\rm H}$ and $R_{\rm 23.5,i}$. However, for $R_{\rm e}$ and $R_{\rm e,M_{\star}}$, the scatter due to the background correction is negligible.

The other significant source of scatter in the size - mass plane is the uncertainty in measuring the total stellar mass of  galaxies from the integrated stellar mass density profile of the objects. As explained in Sect. \ref{sec:parameters}, we measure our total stellar mass by integrating the stellar mass density profiles. To quantify how the uncertainty in the total stellar mass affects our results, we have assumed the following uncertainties in measuring the stellar mass: $\delta_{mass}$=0.24$\pm$0.01 dex (for the entire sample), $\delta_{mass}$=0.19$\pm$0.01 dex (for the E0-S0+ subsample), $\delta_{mass}$=0.24$\pm$0.01 dex (for the S0/a-Sm subsample) and $\delta_{mass}$=0.25$\pm$0.03 dex (for the Dwarfs subsample). These values were computed by an analysis of the differences between the Portsmouth stellar masses of our galaxies \citep[][]{2013MNRAS.435.2764M} and those we measured  using the \textsl{g-r} colour profile \citep[][see Appendix \ref{app:masscomparison} for further details]{2015MNRAS.452.3209R}. To model the effect of the mass uncertainty ($\sigma_{\rm R_{mass}}$) on the scatter of the scaling relationship,  all the observed stellar mass profiles were either  scaled  up or down in mass to place the galaxies on the best fit line through the observed stellar mass plane. This has been performed self-consistently, i.e. taking into account the change in the location of $R_{\rm 1}$ due to the scaling of the profile. Once all the galaxies are located exactly on top of the best-fit stellar mass - size relation (i.e. with zero scatter), we randomly scale the stellar mass density profiles up or down  again, this time by a quantity compatible with a Gaussian distribution whose standard deviation is given by the above $\delta_{mass}$ values.  We repeat this procedure 1000 times and on each occasion we measure the scatter of the stellar mass - size plane produced by the uncertainty in measuring the stellar mass. We show an illustration of the  scatter of the  stellar mass - size relation caused by the uncertainty in  stellar mass in Fig. \ref{fig:masstracks}. The scatter in the stellar mass - size plane generated by the uncertainty in  mass is shown in Table \ref{table:fitting}. Interestingly, for $R_{\rm 1}$, $R_{\rm H}$ and $R_{\rm 23.5,i}$, we find that the dwarfs  are the most affected by the uncertainty due to our mass determination. This is once again expected as the star formation activity of dwarf galaxies is, on average,  more stochastic \citep[][]{2014MNRAS.441.2717K} and complicated to model than that of massive spirals and ellipticals. Therefore, a single colour is not  a good proxy for  the M/L ratio of dwarfs as it is in the case for more gentle star formation histories.

Once the scatter produced by both   the uncertainty in the background and the stellar mass determination have been characterised, we can calculate the intrinsic scatter of the stellar mass-size relations. To do this, we  have  taken the observed scatter and   removed in quadrature the two scatters generated by the background level and stellar mass uncertainty. Obviously, the exact intrinsic scatter of the mass - size relation is difficult to measure as there is some ambiguity in choosing the uncertainty in  stellar mass. Here we have used the above uncertainty values in the stellar mass motivated by what we find comparing the Portsmouth stellar masses \citep{2013MNRAS.435.2764M} with the ones we retrieve using the \textsl{g}-\textsl{r} colour \citep{2015MNRAS.452.3209R}. We acknowledge that our intrinsic scatter values are an approximation, but do estimate that the intrinsic scatter of the $R_{\rm 1}$ - mass relation is about a factor of 1.5 smaller than the observed one (i.e. $\sim$0.06 dex), as a crude evaluation. This implies that the intrinsic stellar mass - $R_{\rm 1}$ relation is, indeed, very tight. In future work, we will address this issue in much more detail. This will be possible as a result of deeper data and therefore a decrease in the uncertainty in measuring the image background level. In addition, we plan to explore the stellar mass - $R_{\rm 1}$ relation using  3.6$\mu$m images (S. D\'iaz-Garc\'ia et al, in prep.) from Spitzer, where the uncertainty in measuring the stellar mass is smaller. In particular, we will use the S$^4$G survey \citep[][]{2010PASP..122.1397S} analysis where images have been corrected by the contamination from young stars \citep[][]{2015ApJS..219....5Q} and the depth is enough to reach 1 $M_{\rm \odot}$ pc$^{-2}$ \citep[][]{2015ApJS..219....3M}.

\begin{figure}
\includegraphics[width=1.0\linewidth]{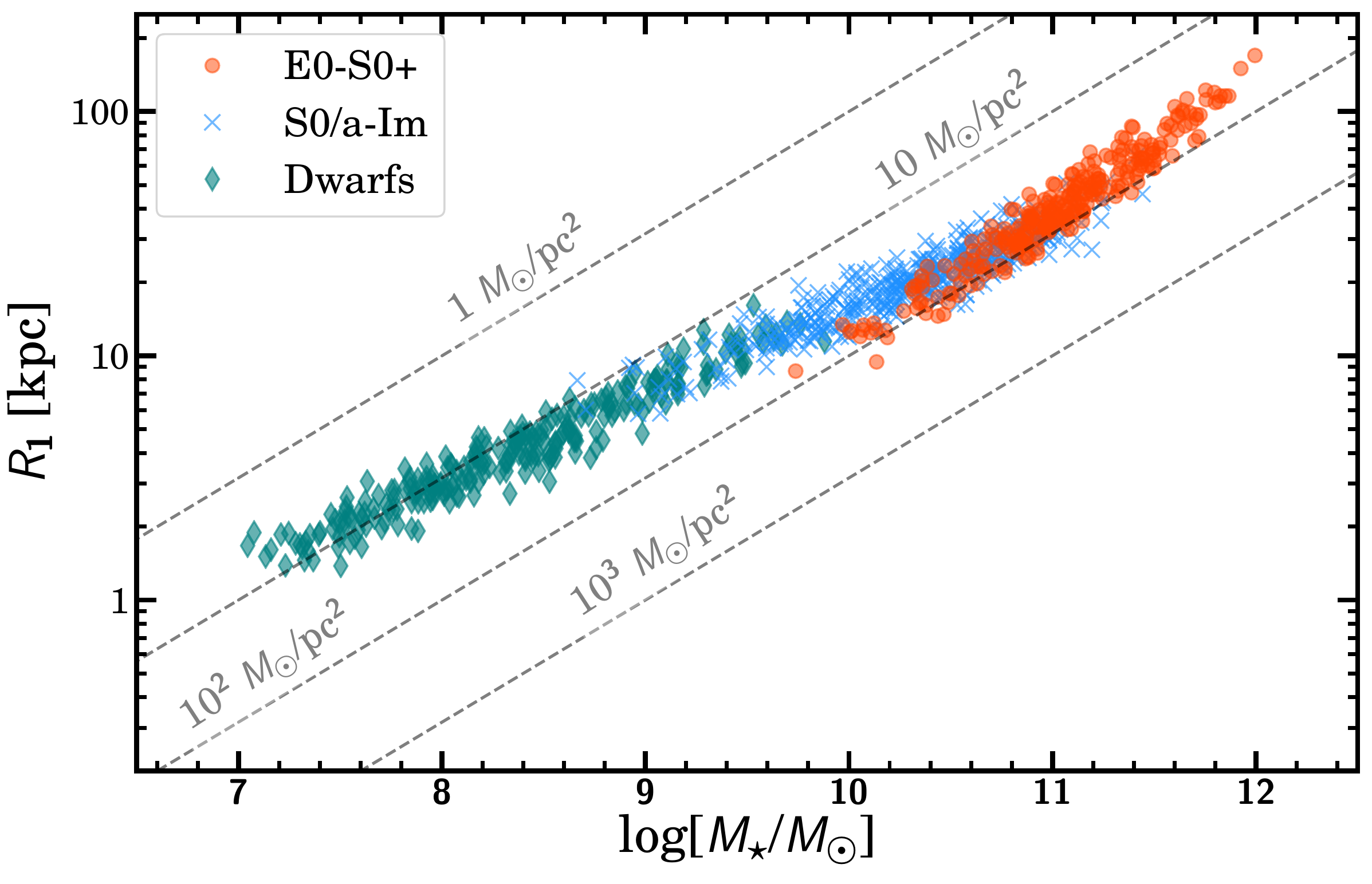}
\caption{Similar to Fig. \ref{fig:scale_relation} (bottom panel), this figure shows the $R_{\rm 1}$--stellar mass relation as it would be observed without any uncertainty in measuring the background level of the images and the stellar mass of  galaxies. The intrinsic scatter of the relation (0.06 dex) is a factor of 2.5 smaller than the scatter of the $R_{\rm e}$ - mass relation.}
\label{fig:intrinsic_relation}
\end{figure}

Finally, it is worth mentioning how the intrinsic scatter of the new mass - size relation compares with the intrinsic scatter of the other popular size - mass relations. In the case of $R_{\rm e}$ and $R_{\rm e,M_{\star}}$, the uncertainty produced by an incorrect background determination is almost negligible. We have found for these cases that $\sigma_{R_{back}}\sim$0.001 dex. This is because our images are very deep and therefore the effect of the uncertainty in our background estimation on the surface brightness profiles barely affects the location of $R_{\rm e}$ and $R_{\rm e,M_{\star}}$ which are found at relatively high surface brightness values. Therefore, we do not expect a large contribution to the observed scatter of these size - mass relations from incorrect background level measurements. In the case of the uncertainty in stellar mass, the first thing to note is given that $R_{\rm e}$ ($R_{\rm e,M_{\star}}$) is defined as the location where half of the total light (stellar mass) is enclosed, its measurement is not affected by an incorrect mass determination of the object. This is because the only effect any uncertainty in mass could transfer to the shape of the profile is a scaling factor towards higher or lower stellar density. However, although the scatter in the size axis is negligible, the uncertainty in the mass axis will play a role in the total scatter of the size - mass plane. Nonetheless, there are two reasons why such an uncertainty will play a minor role in the observed scatter of these relations. Firstly, the slope of the $R_{\rm e}$ ($R_{\rm e,M_{\star}}$) - mass relation is rather flat in the region 10$^9$ to 10$^{11}$ $M_{\rm \odot}$, therefore the contribution from an uncertainty in the mass to enlarge the scatter of the relationships in this region would be close to zero (for example, in spirals and $R_{\rm e}$, the observed scatter is  $\sigma_{\rm R_e}$=0.162$\pm$0.009 while the intrinsic scatter is almost the same $\sigma_{\rm R_{e,int}}$=0.155$\pm$0.009). Secondly, as the observed scatters of the $R_{\rm e}$  ($R_{\rm e,M_{\star}}$) - mass relations are already  larger than in the case of $R_{\rm 1}$ ($\sigma_{\rm R_{obs}}$), the contribution of a similar uncertainty in mass to the intrinsic scatter is very small. In the case of the $R_{\rm e}$-mass relation, as can be seen in Table \ref{table:fitting}, the global observed scatter is only reduced by 9\% after accounting for the mass uncertainty, giving an intrinsic scatter of $\sigma_{\rm R_{e,int}}$=0.154$\pm$0.009. Therefore, it is reasonable to compare the scatter of the \textit{observed} $R_{\rm e}$-mass relation with the \textit{intrinsic} $R_{\rm 1}$-mass relation. As shown in Table \ref{table:fitting}, the decrease in  scatter from $R_{\rm 1}$ to $R_{\rm e}$ ranges from a factor of 2.5 (comparing both intrinsic scatters) to 2.75 (comparing the intrinsic scatter using $R_{\rm 1}$ with the observed scatter using $R_{\rm e}$). We illustrate in Fig. \ref{fig:intrinsic_relation} how the $R_{\rm 1}$ - mass relation would be observed without the scatter produced by the background level and the stellar mass determination. Finally, for the isophotal sizes $R_{\rm H}$ and $R_{\rm 23.5,i}$,  the main contributor to the observed scatter is also the uncertainty in measuring the global mass of galaxies.
 In these cases, the intrinsic scatter for the global size - mass relation decreases by 15-20\% compared to the observed values. Compared to the $R_{\rm 1}$ - mass relation, the intrinsic scatter of the global size - mass relations using $R_{\rm H}$ and $R_{\rm 23.5,i}$ is a factor of 1.5 and 1.4 larger, respectively. In 
 Sec. \ref{subsection:scatters}, we expand 
 on these results by comparing the scatter of the size - mass relations as a function of  galaxy morphology.

\section{Discussion}  
\label{section:discussion}

The results of this paper show that the use of a physically motivated definition for the size of  galaxies based on the location of the gas density threshold for star formation produces a global stellar mass - size relation with a very narrow intrinsic scatter (0.06 dex). In the following subsections, we compare the characteristics of the new size parameter $R_{\rm 1}$ as well as the $R_{\rm 1}$ - mass relation with other popular size measurements.

\subsection{$R_{\rm 1}$ compared to other popular size definitions}
\label{subsection:scatters}

\begin{figure}
\includegraphics[width=\linewidth]{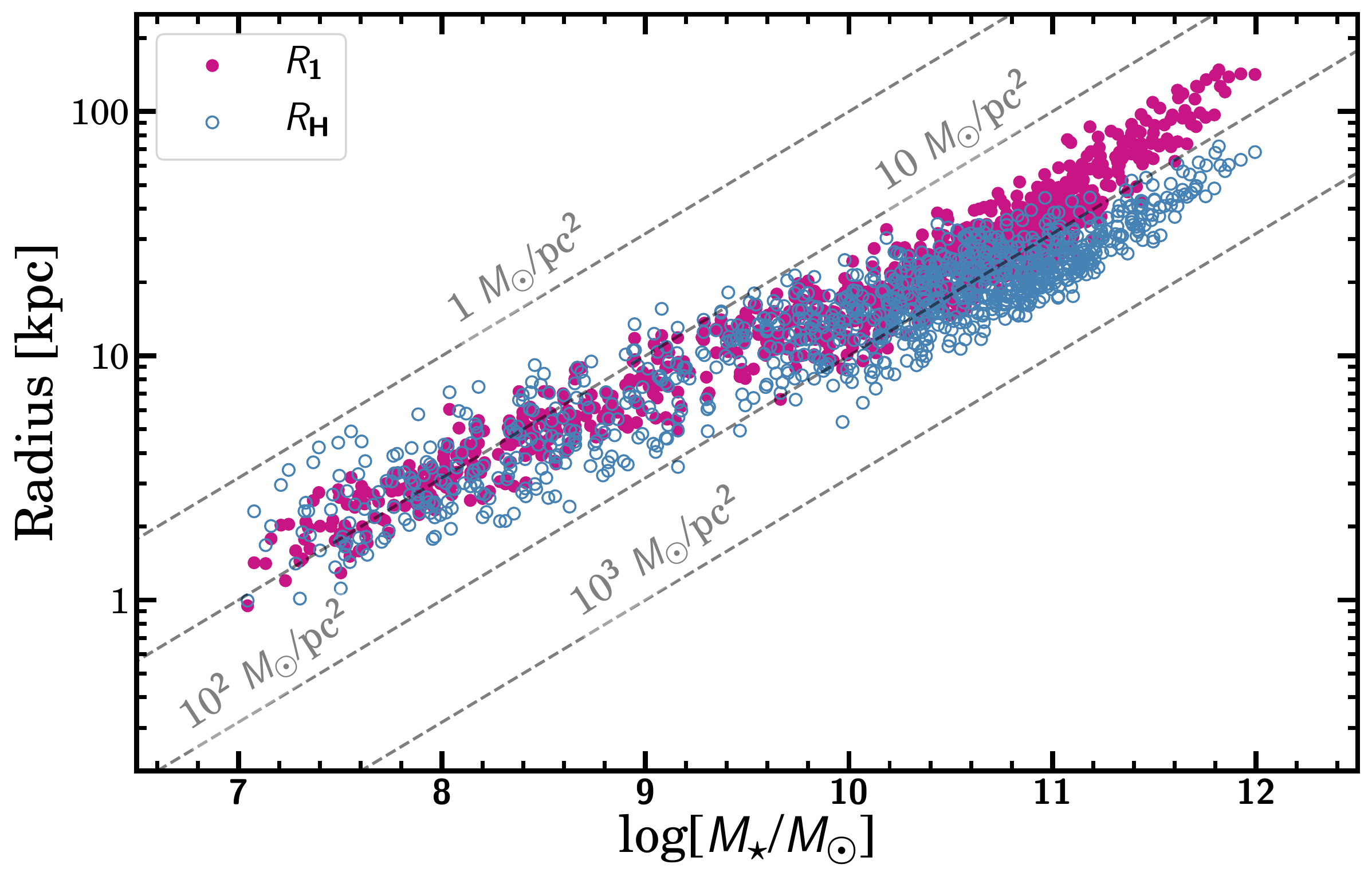}
\includegraphics[width=\linewidth]{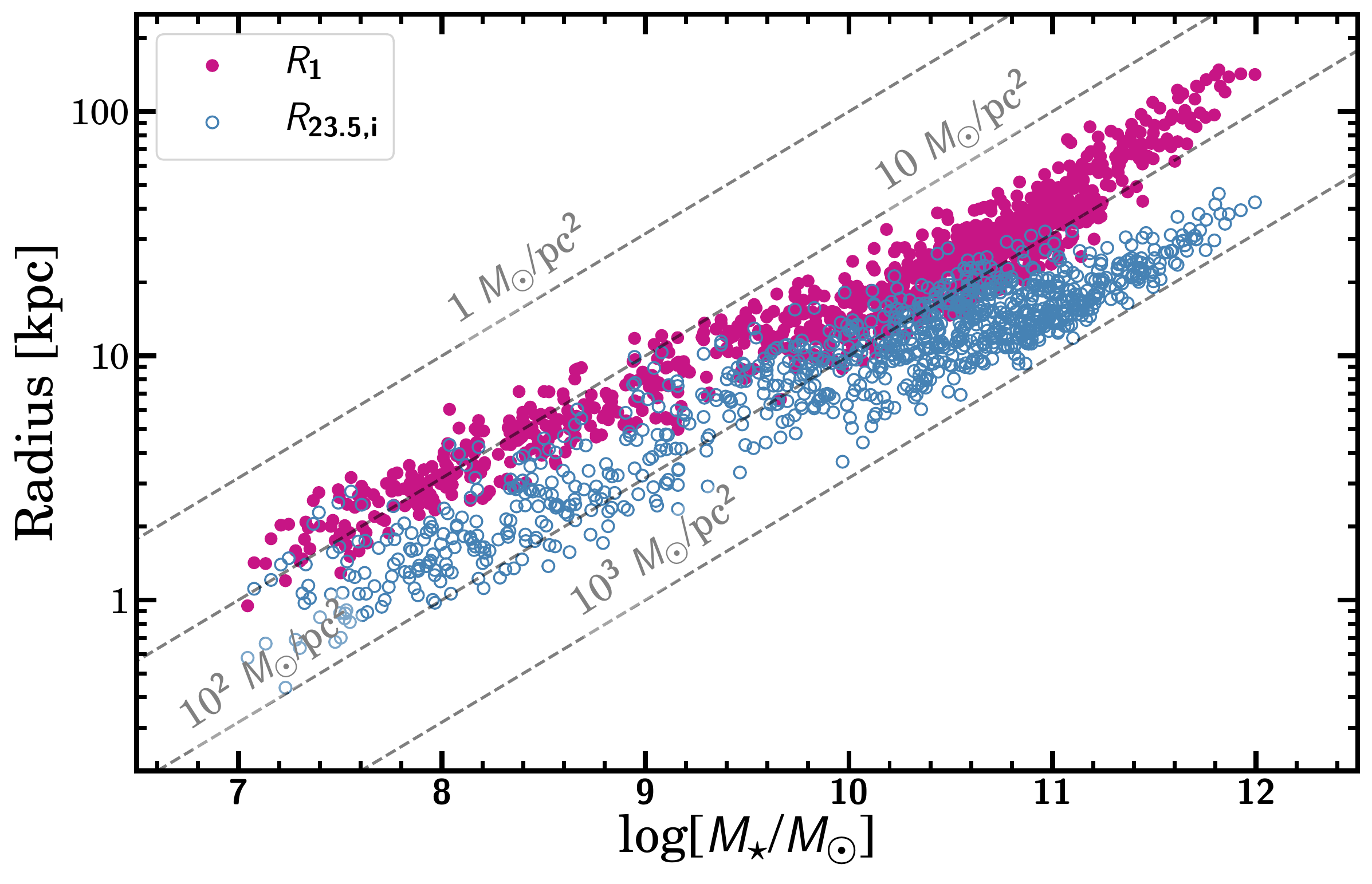}
\includegraphics[width=\linewidth]{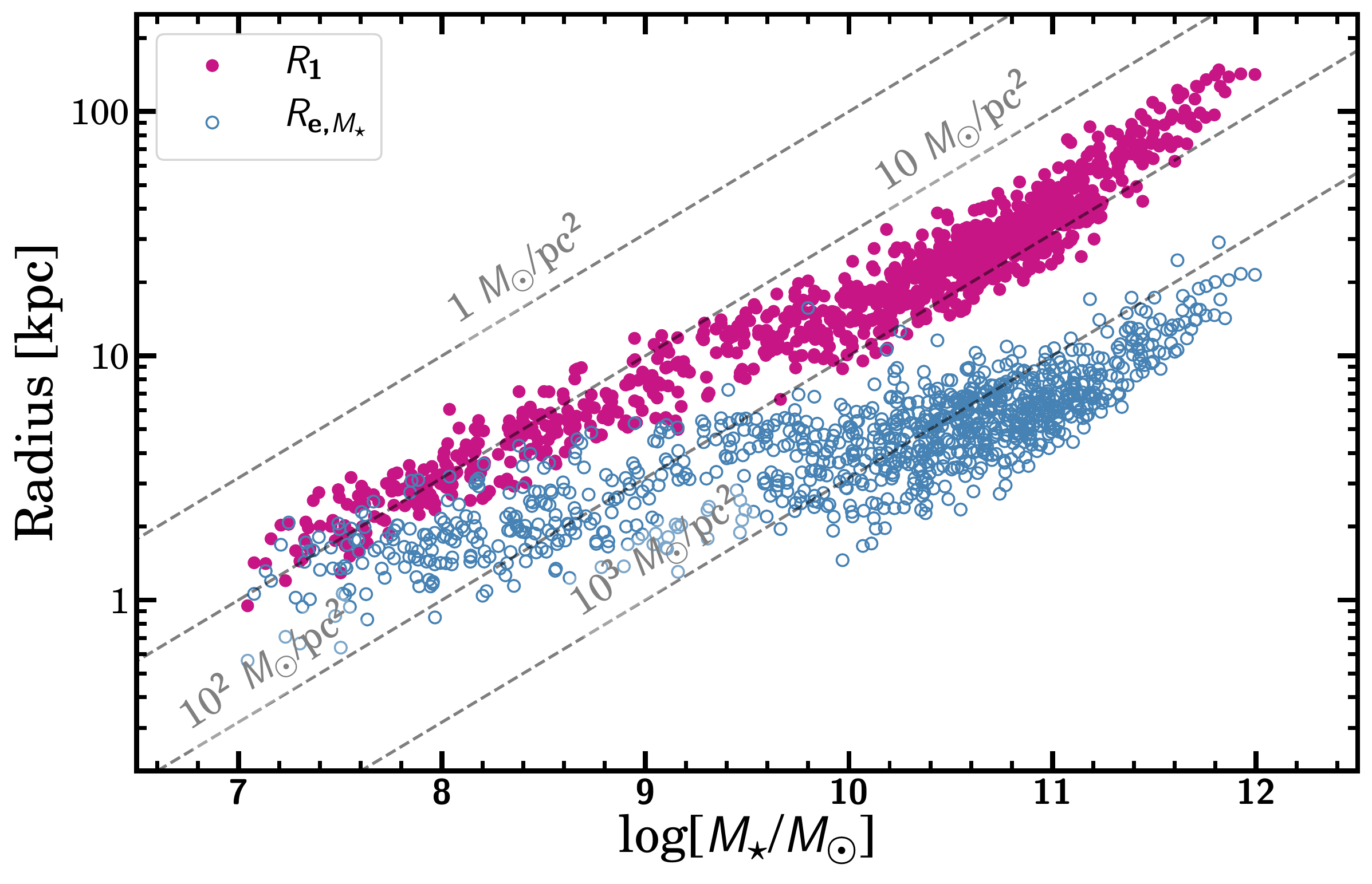}
\caption{Comparing the $R_{\rm 1}$ - mass relation with other size - mass relations using:  Holmberg Radius, $R_{\rm H}$ defined in this work as the 26 mag/arcsec$^{-2}$ isophote in SDSS \textsl{g}-band (upper panel),  $R_{\rm 23.5,i}$,    the radial location of the  $\mu_i$=23.5 mag/arcsec$^2$ isophote (middle panel) and $R_{\rm e,M_{\star}}$, the half mass radius (lower panel).}
\label{fig:holmberg}
\end{figure}

In this paper we have used $R_{\rm 1}$ as a proxy for the location of the  gas density threshold for star formation in galaxies. Nonetheless, the use of $R_{\rm 1}$ as a size indicator is reminiscent of definitions based on the B-band isophote at 25 mag/arcsec$^2$, at 26.5 mag/arcsec$^2$ (i.e. the Holmberg radius)  or in the \textsl{i}-band such as $R_{\rm 23.5,i}$. Although our size definition is not based on the depth of current surveys (as was the case for the size parameters that were defined using photographic plates), it is worth exploring the stellar mass - size plane with popular isophotal size definitions. In this work we lack the B-band filter. As a compromise, we thus decided to use \textsl{g}-band imaging (the closest filter to the B-band with enough depth)  available to us to show the stellar mass - size plane when using the position of the 26 mag/arcsec$^2$ (\textsl{g}-band) isophote as a size indicator. It is this isophote to which we refer to as the Holmberg radius ($R_{\rm H}$). We also include a comparison with  size based on the location of the \textsl{i}-band isophote 23.5 mag/arcsec$^2$ ($R_{\rm 23.5,i}$). The results of this exercise are shown in Fig. \ref{fig:holmberg}.

The observed scatter of the global stellar mass - $R_{\rm H}$ relation is 0.109 $\pm$ 0.005. This value is larger than the one observed for $R_{\rm 1}$ (0.089$\pm$0.005). Interestingly, the scatter is particularly larger for the dwarfs and spirals than for the massive ellipticals. This is understandable as the variability in star formation activity among the less massive galaxies is larger than for the most massive ellipticals. Different star formation levels produce different $g$-band luminosities for the same stellar mass density, and therefore the scatter is larger when using size indicators based on blue bands (as is the case of $R_{\rm H}$). A potential way to decrease the scatter using a single photometric band would be to use a redder band (i.e. one less affected by  recent star formation activity). For instance, one would expect the use of the \textsl{i}-band to decrease the scatter of the stellar mass - size relation. This is in fact the case. The observed scatter of the global stellar mass - $R_{\rm 23.5,i}$ relation is a bit lower (0.106 $\pm$ 0.006 dex) than in the case of $R_{\rm H}$ using the \textsl{g}-band.

While the observed scatter of the $R_{\rm 1}$-mass relation is predominantly affected  by the uncertainty in background and mass estimation, in the case of the $R_{\rm H}$-mass and $R_{\rm 23.5,i}$-mass relations, the main contributor to the scatter is the mass uncertainty. This is because the 26 mag/arcsec$^2$ isophote in the \textsl{g}-band and the 23.5 mag/arcsec$^2$ in the \textsl{i}-band are brighter than the typical brightness of the isomass contour 1 $M_{\rm \odot}$ pc$^{-2}$ (see Sec. \ref{sec:intrinsic}). Consequently, the contribution of the uncertainty in the background to the estimation of the location of $R_{\rm H}$ (and $R_{\rm 23.5,i}$) is not very important. Therefore, while the intrinsic scatter of the $R_{\rm 1}$-mass relation is around 0.06 dex, for $R_{\rm H}$-mass (and $R_{\rm 23.5,i}$ - mass) the global intrinsic scatter  decreases to $\sim$0.09 dex.  We  discuss our findings in the context of the H{\sc i} - mass relation of galaxies in Appendix \ref{app:hi}.

Although the global size - mass relation using  $R_{\rm 1}$  produces the smallest scatter,  it is worth checking whether this is also the case for  different galaxy families. The family of galaxies that consistently shows both the lowest observed and intrinsic  scatter values in the stellar mass - size plane is the E0-S0+ group. This applies to all the size indicators explored, including  the effective and half-mass radii. It is particularly remarkable that the observed scatter using $R_{\rm 23.5,i}$ (0.056$\pm$0.005 dex) is almost comparable to the lowest intrinsic scatter values obtained for this galaxy type  using $R_{\rm 1}$ and the half-mass radius ($\sim$0.04 dex). The small scatter of the elliptical galaxies is a direct consequence of their low level of internal structure compared to other galaxies. This fact makes the members of this family  almost homologous. Therefore, if one is interested in a relative comparison between the size of  galaxies within such a family,  any size indicator already suggested in the literature is useful.

In the case of the S0/a-Sm family, i.e. those galaxies with very complex internal structure consisting of bars, rings, spiral arms, etc, the difference in scatter among the  size indicators is much larger than for the ellipticals. As expected, the  size indicators showing the larger scatter for this galaxy type are those which better reflect the light concentration of the objects: i.e. the effective and the half-mass radii. However, those size measurements that are closer to a characterisation of the boundaries  of the galaxies (e.g. $R_{\rm 1}$, $R_{\rm H}$ and $R_{\rm 23.5,i}$) are the ones with lower scatter. A similar result is found for the dwarf galaxies.

\begin{figure}
\includegraphics[width=\linewidth]{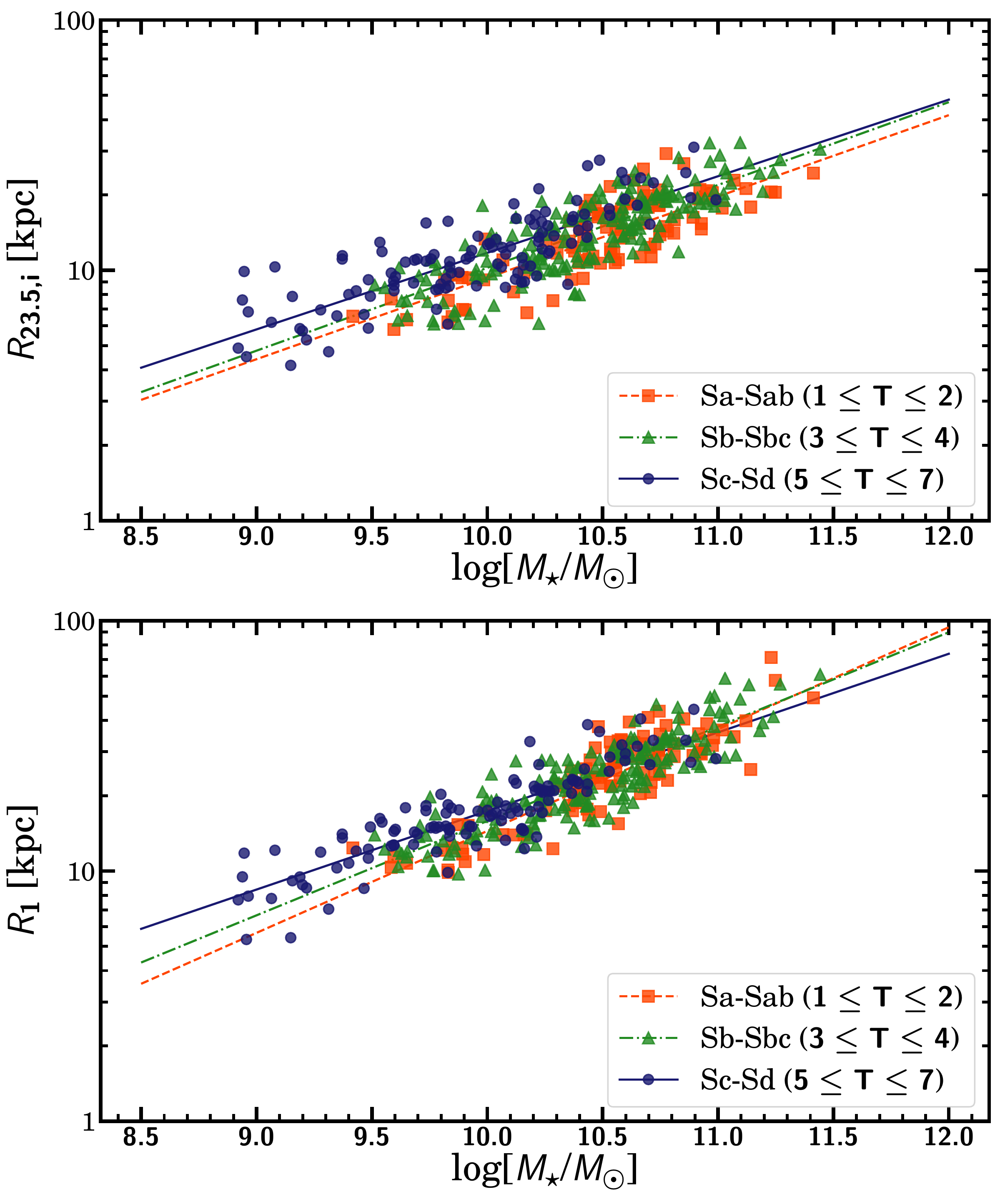}

\caption{Stellar mass - size relation for three morphological groups within our spiral galaxy sample: Sa-Sab (orange squares), Sb-Sbc (green triangles) and Sc-Sd (blue dots). The top panel shows the relation using $R_{\rm 23.5,i}$ while the bottom panel shows the same relation using $R_{\rm 1}$ as the size indicator. Similar to \citet{2011ApJ...726...77S},   \citet{2012MNRAS.425.2741H} and \citet{2015ApJS..219....3M}, we find that spiral galaxies are stratified, with the largest ones at  fixed stellar mass being those of later morphological types.}
\label{fig:spirals}
\end{figure}

In addition to the above results, we also quantitatively compare the scatters found here for spiral galaxies with  those measured in the literature. Similar to \citet{2011ApJ...726...77S} and \cite{2012MNRAS.425.2741H}, we divide our spiral galaxy sample into three categories: Sa-Sab, Sb-Sbc and Sc-Sd. In Fig. \ref{fig:spirals} we show the stellar mass - size relations for these types using $R_{\rm 1}$ and R$_{23.5i}$ as size indicators. We find a similar stratification as the one reported by \citet{2011ApJ...726...77S},  \citet{2012MNRAS.425.2741H} and \citet{2015ApJS..219....3M}, i.e. at fixed stellar mass (or luminosity), those galaxies having later types are the largest. This is especially manifested at the low mass end. \citet{2011ApJ...726...77S} has a sample mostly composed of Sc galaxies. Using R$_{23.5i}$ and the luminosity in the \textsl{i}-band, they  found an observed scatter of 0.05 dex. For the same morphological type, we find here (this time using the stellar mass) an observed scatter of 0.101$\pm$0.007 dex. The larger scatter is connected to the fact that we use the stellar mass instead of the luminosity. The observed scatter using $R_{\rm 1}$ for Sc-Sd galaxies is 0.082$\pm$0.007 dex. As our main source of the scatter is the determination of the stellar mass, it is worth giving the intrinsic scatter values: 0.096$\pm$0.007 dex (R$_{23.5i}$) and 0.066$\pm$0.007 dex ($R_{\rm 1}$). Within the common mass range 10$^{9.5}$-10$^{11}$ $M_{\rm \odot}$ for all the spiral galaxy types, the observed scatters for the Sc-Sd galaxies are: 0.095$\pm$0.008 dex (R$_{23.5i}$) and 0.077$\pm$0.008 dex ($R_{\rm 1}$). The scatter reported by  \citet{2011ApJ...726...77S} is extraordinarily tight. Using a similar sample, \citet{2012MNRAS.425.2741H} found an observed scatter value for the R$_{23.5i}$-mass relation which ranges from 0.070 dex (for their higher quality sample) to 0.096 dex (their entire sample), which is in closer agreement with our observed value.

Another aspect to highlight is the change in the global slope of the stellar mass - size relation as a function of the size indicators we have explored.  Using $R_{\rm 1}$, we find a slope a  bit above  1/3 between 10$^7$ to 10$^{11}$ $M_{\rm \odot}$. This value is in line with the one found  using isophotal radii ($R_{\rm H}$ abd $R_{\rm 23.5,i}$) as a size measure. The slopes, however, decrease significantly when using the effective and half-mass radii. We expand on the potential meaning of the slope we measure using $R_{\rm 1}$ in subsection \ref{slopesection}.  

\subsection{The slope of the stellar mass - size relation}
\label{slopesection}

The slope of the stellar mass - size relation we report for galaxies within the mass range 10$^7$ to 10$^{11}$ $M_{\rm \odot}$ is very close to 1/3. A straightforward calculation shows that if all the stars within $R_{\rm 1}$ were located within a sphere of such radius, the stellar mass density (in 3D) of all the galaxies in this mass range will be  equivalent to $\sim$4.5$\times$10$^{-3}$ $M_{\rm \odot}$ pc$^{-3}$. Obviously, the spatial configuration of both dwarf and spiral galaxies is not spherical but disc-like. Nonetheless, it is suggestive to think that the gas that originally formed all these objects was in a spherical-like configuration at an early galaxy phase before its collapse to form the disc configuration. In other words, it is worth speculating whether  the currently observed  3D stellar density for all the galaxies in our sample is a reflection of a common 3D gas density at an early phase  (before collapsing) of our objects. In fact, this constant 3D stellar density could be linked with the expected constant density of dark matter haloes which formed at a given age of the Universe \citep[see e.g.][]{1998MNRAS.295..319M}. 

It is also worth indicating that while we see a monotonic increase in the size of galaxies with stellar mass using $R_{\rm 1}$ (as well as for $R_{\rm H}$ and $R_{\rm 23.5,i}$), the same is not true for $R_{\rm e}$ or  $R_{\rm e,M_{\star}}$. This is particularly manifested in the interval 10$^9$ to 10$^{11}$ $M_{\rm \odot}$ in stellar mass where the increase in effective (or in the half-mass) radius of the (mostly) spiral galaxies is very modest. This mass range is  where the bulges of  spiral galaxies appear. What we are witnessing here is the enormous impact of using $R_{\rm e}$ (or  $R_{\rm e,M_{\star}}$) for measuring  sizes when a significant amount of the light (or stellar mass) of  galaxies can be concentrated in the inner parts of  galaxies with a bulge. This small increase in $R_{\rm e}$ (or  $R_{\rm e,M_{\star}}$) between 10$^9$ to 10$^{11}$ $M_{\rm \odot}$ is not a minor issue. As the vast majority of works aiming to understand the connection between the galaxy size and the dark matter halo properties use $R_{\rm e}$ as a size indicator \citep[see e.g.][]{2013ApJ...764L..31K,2019MNRAS.tmp.1977J,2020MNRAS.492.1671Z}, the small increase in the effective radius in this mass range can hide a potential connection between the dark and the luminous component of the galaxies. In a companion paper (C. Dalla Vecchia et al. in prep.), we show that the use of $R_{\rm 1}$ permits the connection of both galaxy components directly, ultimately facilitating our understanding about how these objects form. 

\subsection{The tilt of the stellar mass - size relation at 10$^{11}$ $M_{\rm \odot}$}
\label{subsec:tilt}

A notable feature  of the new stellar mass - size relation is the change in  slope observed at $\sim$10$^{11}$ $M_{\rm \odot}$. The slope changes from $\sim$1/3 to $\sim$3/5 (see Table \ref{table:fitting}) for the most massive galaxies. The abrupt change in slope is found in all the size indicators probed in this work. This $\sim$10$^{11}$ $M_{\rm \odot}$ stellar mass value marks the shift between objects with disc-like configuration to objects with a spherical symmetry. In addition, this is the stellar mass where the transition from rotationally to pressure-supported systems has been  reported \citep[see e.g.][]{2011MNRAS.414..888E}.

As mentioned in  Appendix \ref{starformationproxy}, we speculate that this change in  slope is a manifestation of  different gas density threshold values for star formation  in galaxies that formed at high-$z$ could have had. Observational evidence has shown that the most massive galaxies underwent a huge burst of star formation at high-z with star formation rates reaching values $\gtrsim$1000 $M_{\rm \odot}$ yr$^{-1}$ \citep[see e.g.][]{2013Natur.496..329R}. The high star formation rates these galaxies have undergone could have injected a lot of energy into the gas, thereby preventing the star formation at low mass densities and consequently increasing the gas density threshold for star formation. A remnant of this huge star formation burst is the core of these massive galaxies that later undergo important merger activity,  creating their envelopes \citep[see e.g.][]{2011MNRAS.415.3903T,2014MNRAS.444..906F,2017MNRAS.466.4888B}. 

In short, we speculate that the tilt we observe at $\sim$10$^{11}$ $M_{\rm \odot}$ in the new stellar mass - size plane is a reflection of a change in the gas density threshold for star formation when the bulk of the most massive galaxies originated. 

\section{Conclusions}  
\label{section:conclusions}

We  introduce a new approach to define the luminous size of galaxies, aiming to link size with the region where galaxies form stars. In order to make such a physically motivated size definition operative, we  propose using the average radial location of the gas density threshold for star formation to measure this quantity. We  suggest the use of the radial position of the isomass contour at 1 $M_{\rm \odot}$ pc$^{-2}$ (here referred to as $R_{\rm 1}$) as a proxy for measuring this threshold. This value is motivated by both theoretical and observational arguments. In particular, the density value found at the location of the truncation in galaxies similar to our own Milky Way.

When using $R_{\rm 1}$ as a size indicator for galaxies, the global scatter of the stellar mass - size relation explored over five orders of magnitude in stellar mass drops significantly, reaching a value of $\sim$0.06 dex. This value is 2.5 times smaller than the scatter measured using the effective radius ($\sim$0.15 dex) and  1.5 to 1.8 times smaller than those using other traditional sizes indicators such as $R_{\rm 23.5,i}$ ($\sim$0.09 dex), $R_{\rm H}$ ($\sim$0.09 dex) and $R_{\rm e,M_{\star}}$ ($\sim$0.11 dex). 

Between 10$^7$ and 10$^{11}$ $M_{\rm \odot}$, the slope of the stellar mass - size relation is very close to 1/3. In a 3D spherical distribution, this corresponds to a constant stellar density of $\sim$4.5$\times$10$^{-3}$ $M_{\rm \odot}$ pc$^{-3}$, which could be a reflection of a common gas density when the primordial gas collapsed to form stars. Beyond 10$^{11}$ $M_{\rm \odot}$, the stellar mass - size relation gets steeper, reaching a slope of $\sim$3/5. We speculate that this drastic increase in  size of the most massive galaxies could be linked to its different star formation histories, reflecting that the gas density threshold for star formation was higher at the epoch of their main formation burst.

\section*{Acknowledgements}

We acknowledge the referee for a careful reading of this paper and a for a large number of suggestions to improve its presentation. We thank  Ra{\'u}l Infante-Sainz and  Javier Rom{\'a}n for providing the extended SDSS Point Spread Functions (PSFs) of all filters for use in this work and St\'ephane Courteau for interesting comments. NC thanks Caroline Haigh for providing the latest version of \texttt{MTObjects}.

We acknowledge financial support from the European Union's Horizon 2020 research and innovation programme under Marie Sk\l odowska-Curie grant agreement No 721463 to the SUNDIAL ITN network, from the State Research Agency (AEI) of the Spanish Ministry of Science, Innovation and Universities (MCIU) and the European Regional Development Fund (FEDER) under the grants with reference AYA2016-76219-P and AYA2016-77237-C3-1-P, from IAC projects P/300624 and P/300724, financed by the Ministry of Science, Innovation and Universities, through the State Budget and by the Canary Islands Department of Economy, Knowledge and Employment, through the Regional Budget of the Autonomous Community, and from the Fundaci\'on BBVA under its 2017 programme of assistance to scientific research groups, for the project "Using machine-learning techniques to drag galaxies from the noise in deep imaging".

 This work has made use of data from the European Space Agency (ESA)
mission {\it Gaia} (\url{http://www.cosmos.esa.int/gaia}), processed by
the {\it Gaia} Data Processing and Analysis Consortium (DPAC,
\url{http://www.cosmos.esa.int/web/gaia/dpac/consortium}). Funding
for the DPAC has been provided by national institutions, in particular
the institutions participating in the {\it Gaia} Multilateral Agreement.

Funding for the Sloan Digital Sky Survey IV has been provided by the Alfred P. Sloan Foundation, the U.S. Department of Energy Office of Science, and the Participating Institutions. SDSS-IV acknowledges
support and resources from the Center for High-Performance Computing at the University of Utah. The SDSS web site is www.sdss.org. SDSS-IV is managed by the Astrophysical Research Consortium for the Participating Institutions of the SDSS Collaboration including the Brazilian Participation Group, the Carnegie Institution for Science, Carnegie Mellon University, the Chilean Participation Group, the French Participation Group, Harvard-Smithsonian Center for Astrophysics, Instituto de Astrof\'isica de Canarias, The Johns Hopkins University, Kavli Institute for the Physics and Mathematics of the Universe (IPMU) / University of Tokyo, the Korean Participation Group, Lawrence Berkeley National Laboratory, Leibniz Institut f\"ur Astrophysik Potsdam (AIP), Max-Planck-Institut f\"ur Astronomie (MPIA Heidelberg), Max-Planck-Institut f\"ur Astrophysik (MPA Garching), Max-Planck-Institut f\"ur Extraterrestrische Physik (MPE), National Astronomical Observatories of China, New Mexico State University, New York University, University of Notre Dame, Observat\'ario Nacional / MCTI, The Ohio State University, Pennsylvania State University, Shanghai Astronomical Observatory, United Kingdom Participation Group, Universidad Nacional Aut\'onoma de M\'exico, University of Arizona, University of Colorado Boulder, University of Oxford, University of Portsmouth, University of Utah, University of Virginia, University of Washington, University of Wisconsin, Vanderbilt University, and Yale University.

This research has made use of the NASA/IPAC Extragalactic Database (NED), which is operated by the Jet Propulsion Laboratory, California Institute of Technology, under contract with the National Aeronautics and Space Administration.

\textit{Software}: \texttt{Astropy},\footnote{http://www.astropy.org} a community-developed core \texttt{Python} package for Astronomy \citep{robitaille2013astropy, price2018astropy}; \texttt{SciPy} \citep{scipy}; \texttt{NumPy} \citep{numpy, doi:10.1109/MCSE.2011.37}; \texttt{Scikit-learn} \citep{scikit-learn}; \texttt{Matplotlib} \citep{Hunter:2007}; \texttt{Jupyter Notebooks} \citep{Kluyver:2016aa}; \texttt{TOPCAT} \citep{2005ASPC..347...29T}; \texttt{IMFIT} \citep{2015ApJ...799..226E}; \texttt{MTObjects} \citep{2016mto}; \texttt{SWarp} \citep{swarp}; and \texttt{SAO Image DS9} \citep{ds9}.




\bibliographystyle{mnras}
\bibliography{bibliography} 


%
%
%

%
%
%


\appendix

\section{Is 1 $M_{\rm \odot}$ pc$^{-2}$ a good proxy for the location of  the gas density threshold for star formation along the  10$^7$ to 10$^{12}$ $M_{\rm \odot}$ mass range?}
\label{starformationproxy}

In this paper we  propose the radial position of the isomass contour  at 1 $M_{\rm \odot}$ pc$^{-2}$ as a size indicator motivated by its proximity to the location of the gas density threshold for star formation in galaxies  found theoretically \citep[see e.g.][]{2004ApJ...609..667S} and because this value is representative of the location of the disc truncation in galaxies similar to the Milky Way \citep[][]{2019MNRAS.483..664M}. However, this value is not guaranteed to be representative of the gas density threshold for star formation in galaxies with very different stellar mass (like  dwarfs) or those who were formed during an enormous burst of star formation at high-$z$ (as is the case of the most massive ellipticals). In fact, there are some hints in our data indicating that a running gas density threshold for star formation could be more representative of the size of  galaxies. Looking at Fig. \ref{fig:examples}, we can appreciate that while  1 $M_{\rm \odot}$ pc$^{-2}$ very well represents the size of  spiral galaxies, the use of this value does not fully enclose the extension of  dwarf galaxies.  A potential explanation for this is that in the case of dwarf galaxies the gas density threshold for star formation  is lower than in the case of more massive spirals, and therefore using, for instance, 0.3-0.5 $M_{\rm \odot}$ pc$^{-2}$ instead of 1 $M_{\rm \odot}$ pc$^{-2}$ could be a better proxy for measuring the size of these small systems. This idea of a lower gas density threshold for star formation for  dwarf galaxies is  observationally supported as on average their amount of H$_2$ compared to H{\sc i}  (and therefore their star-forming efficiency) is lower than for larger mass systems \citep[see e.g.][]{2008AJ....136.2782L,2012ApJ...756..113H}.

\begin{figure}
\centering
\includegraphics[width=0.97\linewidth]{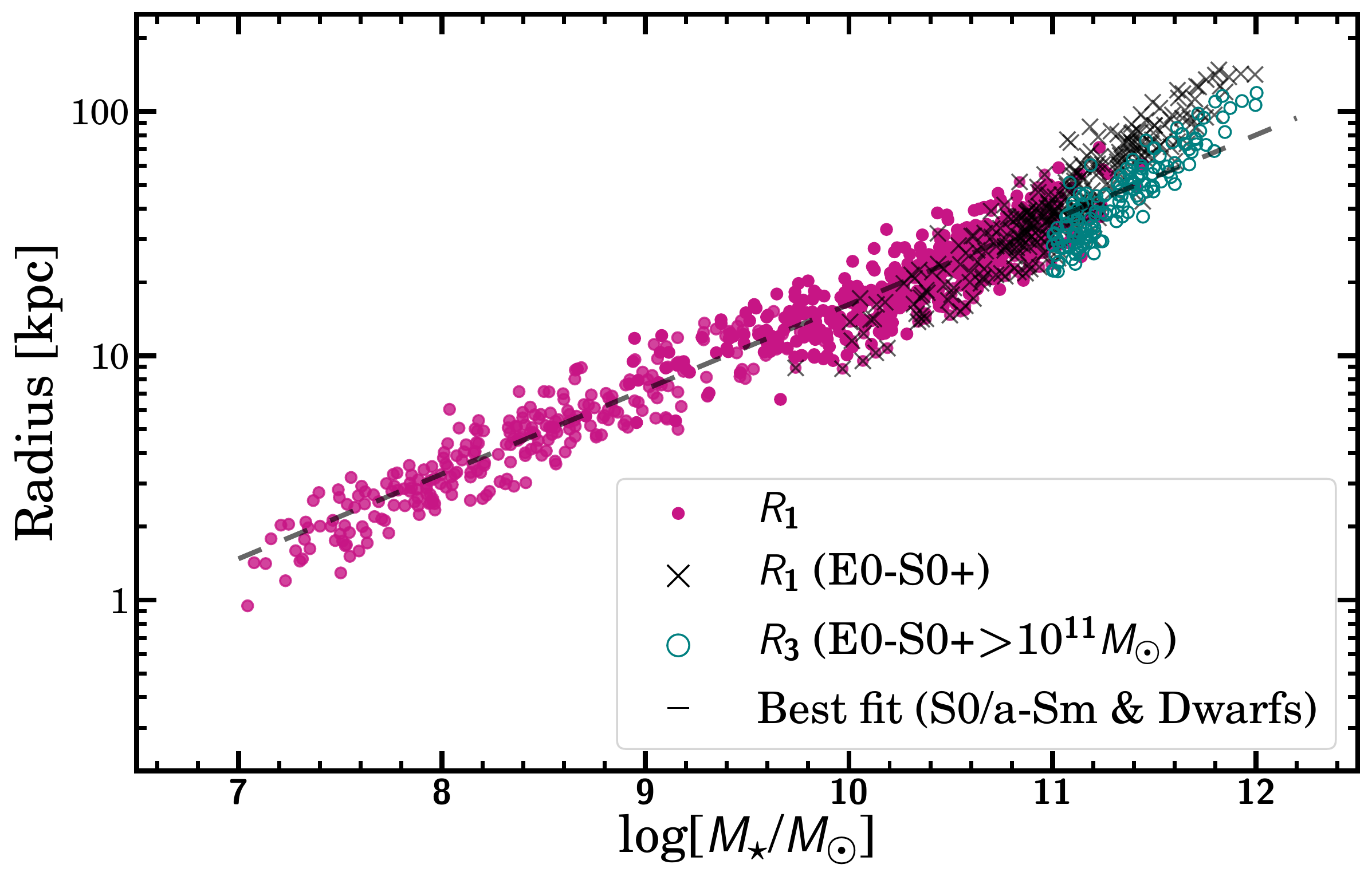} 
\caption{Stellar mass - size relation using different stellar mass densities as proxies for indicating the size of the most massive galaxies: $R_{\rm 1}$ and $R_{\rm 3}$. As can be seen, the size of the most massive objects changes depending on the isomass contour used to measure its size. The figure shows that most of the galaxies from 10$^7$ to 10$^{12}$ could be allocated in the same stellar mass-size relation (i.e. with a similar slope) if $R_{\rm 3}$ were used for measuring the size of the galaxies for objects with stellar mass larger than 10$^{11}$ $M_{\rm \odot}$.} 
\label{fig:slopechange}
\end{figure}

On the other hand, the situation would be reversed in the case of massive elliptical galaxies where $R_{\rm 1}$ is a bit larger than the visual extent of these galaxies. The gas density threshold for star formation during its high-z formation could be larger than whatit is currently for the spiral galaxies. In the local Universe, for instance, it has been found that starburst galaxies have significantly enhanced star formation \citep[][]{2015ApJ...808...66J}. In this case, using  an isomass contour with a larger density value (i.e. $>$1 $M_{\rm \odot}$ pc$^{-2}$) as size could be a better option for these very massive galaxies. In fact,  one of the most notable features of the stellar mass - size relation we have analysed is the change in  slope of the relation for galaxies above 10$^{11}$ $M_{\rm \odot}$ in stellar mass (Sec \ref{subsec:tilt}). We believe that this could be an indication that the proxy we are using to calculate the location of the gas density threshold for star formation for galaxies (i.e. the isomass contour at 1 $M_{\rm \odot}$ pc$^{-2}$) can be slightly different for galaxies where the bulk of star formation occurred at high-$z$. For this reason, we have explored how   the relation changes when we use an isomass contour at 3 $M_{\rm \odot}$ pc$^{-2}$ for the size of the most massive objects. We show the result of doing such an exercise in Fig. \ref{fig:slopechange}.

Using $R_{\rm 3}$ instead of $R_{\rm 1}$ for the most massive galaxies decreases the size of these objects  and thus follows the trend observed in $R_{\rm 1}$ for the less massive galaxies (M$_{\star}<10^{11}$ $M_{\rm \odot}$) as marked by the best fit dashed line in Fig. \ref{fig:slopechange}. Beyond 10$^{11.4}$ $M_{\rm \odot}$, however, even with $R_{\rm 3}$ the galaxies are above this trend in $R_{\rm 1}$. This is suggestive of an even larger gas density threshold for star formation in these ultra massive objects. In \citet{chamba2020}, we describe the opposite exercise using the dwarf galaxies. 

\section{Other potential size indicators based on the location of the gas density threshold for star formation}
\label{app:alternatives}

Intimately linked to the gas density threshold for star formation is the drop-off of galaxy profiles, both in the optical and in H${\sc \alpha}$. The position of a sharp decline in the optical profile, often referred to in the literature as breaks or truncations \citep[see a discussion about this in][]{2012MNRAS.427.1102M}, has been used to explore the cosmic size evolution of galaxies \citep[][]{2005ApJ...630L..17T,2008ApJ...684.1026A}. Unfortunately, not all  galaxies show a clear deviation from the exponential decline in their profiles that can be used to trace their sizes \citep{2006A&A...454..759P}. In this paper, we have used the fact that for Milky Way-like galaxies, the truncation is found near a stellar mass density of $\sim$1 $M_{\rm \odot}$ pc$^{-2}$. However, we do not know whether this result also holds for galaxies of different stellar mass (see Appendix \ref{starformationproxy}). In a future paper, we will explore this   issue with our sample by identifying the location of the truncations in the profiles of our galaxies and determining the values of the stellar mass densities at those positions.

In addition to the drop-off in the optical profile, many disc galaxies show a sharp decline in their H${\sc \alpha}$ profiles. The location of this truncation in H${\sc \alpha}$ is also considered a good indicator of the radial position of the gas density threshold for star formation in actively star forming galaxies \citep[see e.g.][]{2001ApJ...555..301M}. The H${\sc \alpha}$ emission is directly linked to the presence of massive O stars and therefore the presence of recent star formation. In this sense, the location of the truncation radius in H${\sc \alpha}$ could be a direct indicator of the position of the gas density threshold for star formation and consequently, a very good proxy for galaxy size.

Despite the obvious advantages of using H${\sc \alpha}$ profiles to derive the size of  galaxies,  there are a number of practical limitations to a massive use of this technique. First, to the best of our knowledge, there are currently no  deep H${\sc \alpha}$ surveys covering a large area of the sky that allow us to make this analysis on all kinds of galaxies. Second, contrary to the use of a stellar mass isocontour as we have done in this paper, the use of H${\sc \alpha}$ will be restricted to only those galaxies that are currently forming stars. In addition, as is also the case in the optical, not all the disc galaxies show a clear feature in their profiles that can be associated to a sharp decline in on-going star formation \citep[see e.g.][]{2006AJ....131..716K}. Nonetheless, in those cases where the break radii of  H${\sc \alpha}$ surface brightness profiles have been measured, a bimodal distribution has been found, with the majority of the galaxies showing breaks at 0.7 or 1.1 $R_{25}$ \citep{2010MNRAS.405.2549C}. It remains to be explored whether the breaks in H${\sc \alpha}$ are at the same location as that of the truncations in the optical for  large samples of galaxies.

Finally, though the presence of H${\sc \alpha}$ is a clear indication of recent star formation, the H${\sc \alpha}$ line is connected with the presence of stars massive enough  to generate a Stroemgren sphere. However, it is easy to imagine (particularly in the outer parts of  galaxies where the density is very low) star formation occurring at low levels without the formation of very massive stars. In such a case, where H${\sc \alpha}$ is no longer present, the UV emission could be used as an indicator of recent star formation. As mentioned before, it would be necessary to search for a clear feature that can be associated to a gas density threshold for star formation in a large number of galaxies and thus approach the issue of galaxy size and star formation activity from multiple perspectives.

\section{The limits of the new size definition}
\label{app:limits}

Our size definition is motivated by the location of the gas density threshold for star formation in  galaxies. This definition is rather general and there is no reason why this could not be applied to any galaxy where \textit{in-situ} star formation is present. However, in order to make our size definition operative, we have used the position of the stellar mass density contour at 1 $M_{\rm \odot}$ pc$^{-2}$. By taking this specific value, we are limiting ourselves to characterise only the size of objects  whose maximum stellar mass density is $\gtrsim$1 $M_{\rm \odot}$ pc$^{-2}$. 
Therefore, we quantify the brightness of objects that can not be
characterised with such a specific prescription.

Some old and metal-poor galaxies with central surface brightness $\mu_g(0)>$27 mag/arcsec$^2$ would have a maximum stellar mass density $<$1 $M_{\rm \odot}$ pc$^{-2}$ and therefore, they would be unable to have a size estimation based on our proposed isomass contour. As the surface brightness of  galaxies is not usually reported by their central value but as the brightness averaged within their effective radius (i.e. $<\mu>_e$), we  compute  which $<\mu>_e$ limits our size measure $R_{\rm 1}$. This is done by using the relation between central surface brightness and the average surface brightness according to a S\'ersic profile:

\begin{equation}
<\mu>_e=\mu(0)-2.5\log\frac{n\Gamma(2n)}{b_n^{2n}}
\end{equation}
where $n$ is the S\'ersic index of the model. For 0.5$<n<$2 \citep[a typical range of variation of the S\'ersic index for galaxies with very low surface brightness; see e.g.][]{2017MNRAS.468..703R}, this would imply that the use of $R_{\rm 1}$ is limited to galaxies with $<\mu>_e \lesssim$ 28-29 mag/arcsec$^2$. This surface brightness is extremely low, nonetheless the current faintest galaxies detected in our Local Group \citep[see e.g. Fig. 7 in][]{2012AJ....144....4M} have such extreme brightness. This again reinforces the idea we mentioned in Appendix \ref{starformationproxy} that low mass systems have very low star formation efficiency and therefore a lower gas density threshold for star formation. Future ultra-deep imaging surveys would allow us to better calibrate the gas density threshold for star formation values for  low dense systems. Nevertheless, our size definition could be readily  adjusted to these extreme objects by using a lower mass density contour for characterising their sizes.

\section{Stellar mass determination}
\label{app:masscomparison}

As explained in the main body of the paper,  stellar masses of all  galaxies were calculated using their stellar mass density profiles which were derived using the  \textsl{g}-\textsl{r} colour profiles \citep{2015MNRAS.452.3209R}. In order to evaluate the reliability of this mass estimate, we have compared our stellar mass estimations with those obtained by the Portsmouth group \citep[][]{2013MNRAS.435.2764M}. To do that we have retrieved the stellar masses from the Portsmouth Spectro-Photometric Model Fitting catalogue available on the SDSS webpage\footnote{\url{https://data.sdss.org/datamodel/files/BOSS_GALAXY_REDUX/GALAXY_VERSION/portsmouth_stellarmass.html}}. The Portsmouth stellar masses are calculated using the BOSS spectroscopic redshift, $Z_{\rm NOQSO}$ and \textsl{u, g, r, i, z} photometry by means of broad-band SED fitting of stellar population models. Two separate stellar mass calculations were conducted: one assuming a passive template and the other a star-forming template. The version we have used here is the star-forming template with a Kroupa IMF.

\begin{figure}
\centering
\includegraphics[width=\columnwidth]{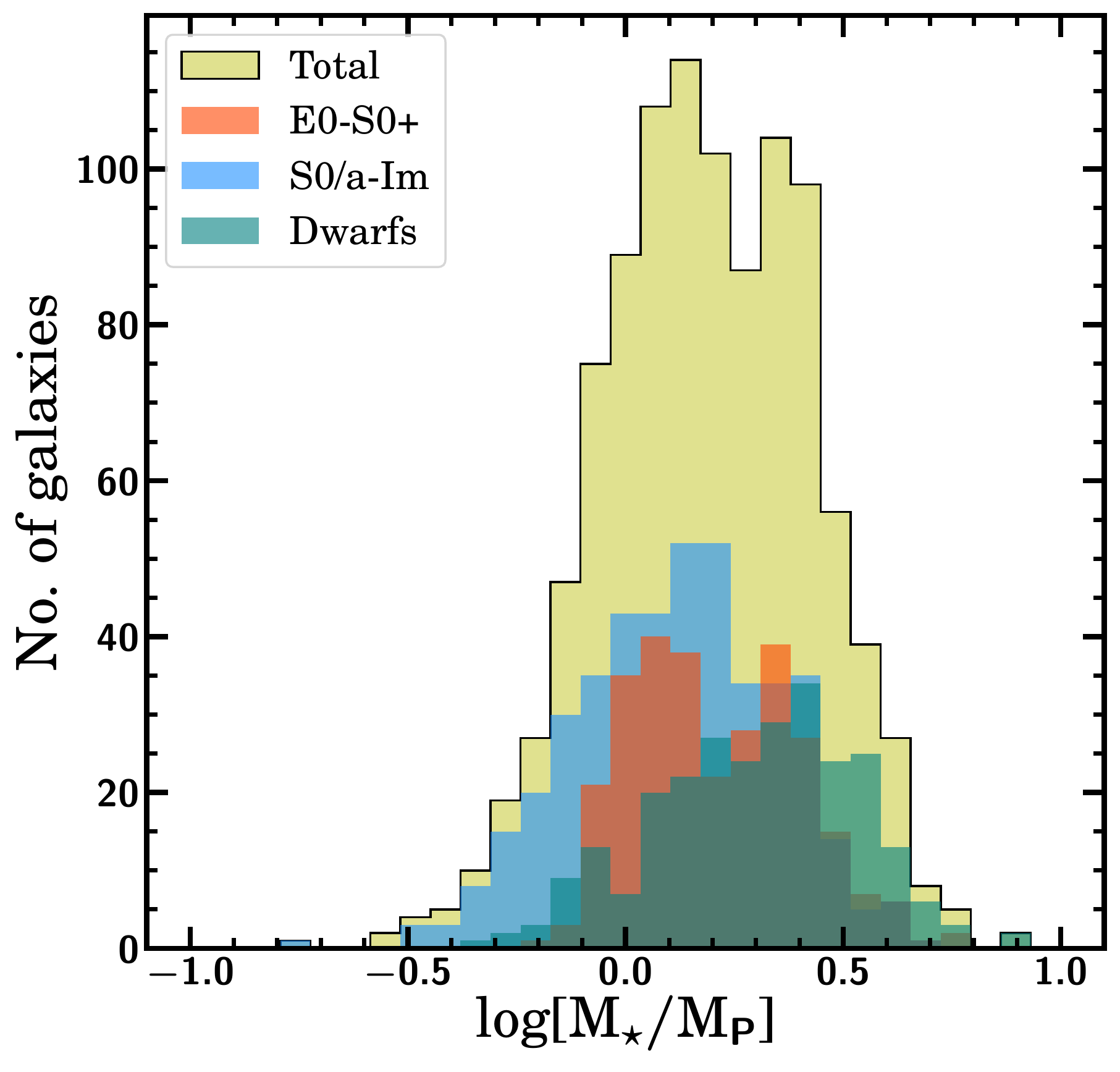}
\caption{Uncertainty in measuring the stellar masses of galaxies used in our sample. $M_{\rm P}$ corresponds to the stellar mass of the objects reported by the Portsmouth group \citep[][]{2013MNRAS.435.2764M} and $M_{\rm \star}$ is the mass we have estimated in this work using the \textsl{g}-\textsl{r} colour profile \citep{2015MNRAS.452.3209R}. We show the ratio of these two estimates for the full sample (yellow) and the different galaxy subsamples labelled in the legend.} 
\label{fig:histogram}
\end{figure}

In Fig. \ref{fig:histogram}, we show the comparison between the stellar masses estimated in this work and those obtained by the Portsmouth group (when available). The ratio of these two mass estimates is approximately described by a Gaussian distribution. 
We find a systematic offset in our mass estimates by a factor of 1.6 with respect to the Portsmouth masses. This offset could be due to multiple reasons - we are not using the same IMF nor the same aperture to estimate the total flux of the objects. However, for our paper, the most relevant aspect is not the offset but the scatter in the difference between the two mass measurements. We calculate this scatter by measuring the r.m.s. of the ratio between the two masses. For the total sample, the r.m.s. of this  distribution is 0.24 dex. In addition to showing the distribution of the ratio  between the two masses of the entire sample, we also overplot the distribution of the same ratio for the different galaxy subsamples defined in this work. We find that the r.m.s. of the subsample distributions are: 0.19 dex (E0-S0+), 0.24 dex (S0/a-Sm) and 0.25 (Dwarfs).

\begin{figure}
\centering
\includegraphics[width=\columnwidth]{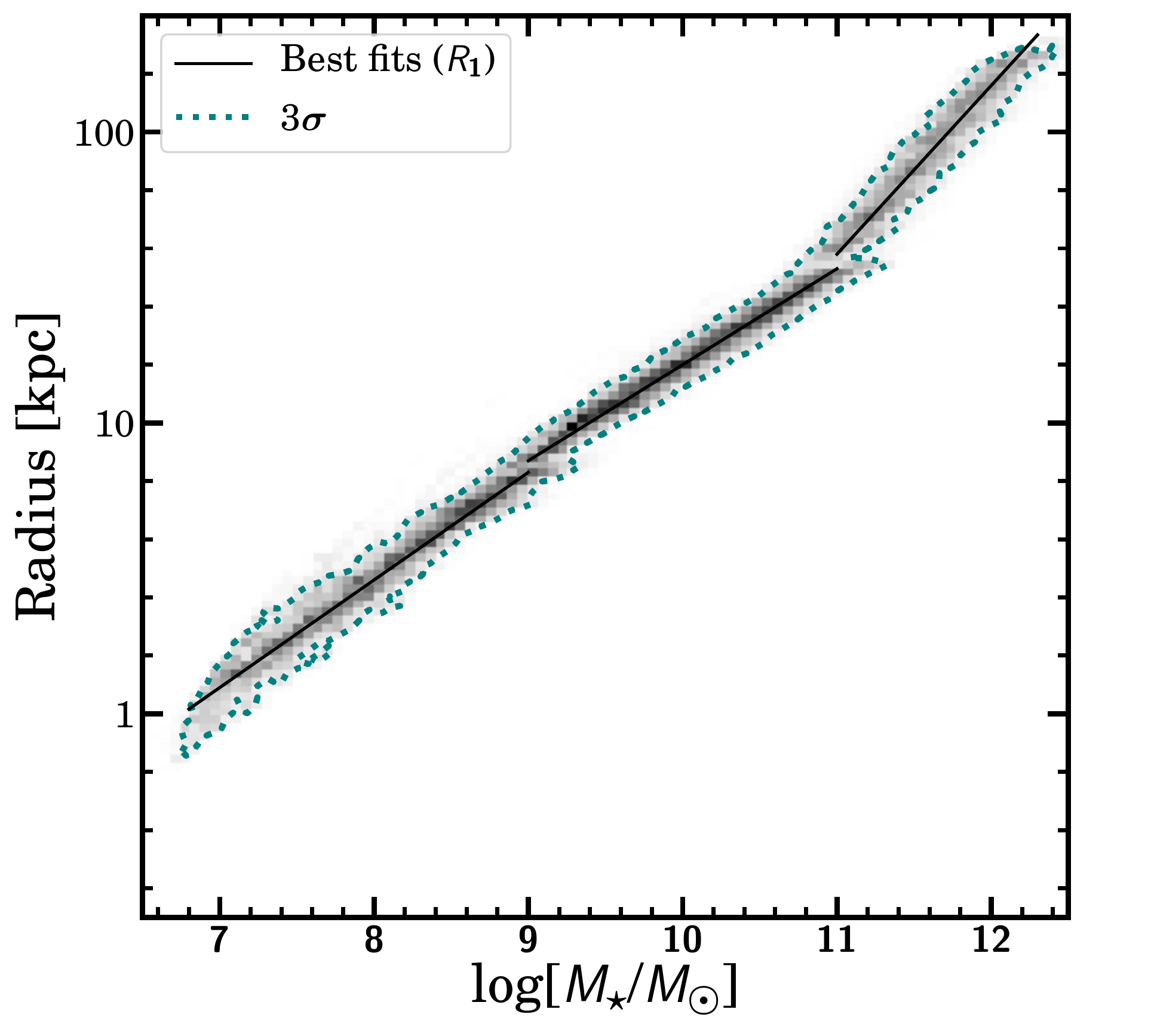}
\caption{Scatter in the $R_{\rm 1}$-mass relation produced by the uncertainty in  stellar mass. The figure shows how  galaxies  spread over the size - mass plane simply due to effect of the uncertainty in measuring the stellar mass of each galaxy. To build this plot we have assumed that the intrinsic scatter of the size - mass relation is exactly zero and follow how the uncertainty in measuring the mass will spread around the best fit lines to the observed measurements. The contour shows the 3 $\sigma$ distribution of the final sample.}  
\label{fig:masstracks}
\end{figure}

In our analysis, we have used the above r.m.s. of the subsample distributions as a way of estimating the uncertainty in the stellar mass - size relation. In Fig. \ref{fig:masstracks}, we illustrate how the $R_{\rm 1}$ - mass relation with an intrinsic scatter equal to zero would look  with only the uncertainty contribution from the stellar mass estimate. In other words, this plot shows the contribution of the mass uncertainty to the observed scatter of the mass - size relation. The scatter of the $R_{\rm 1}$-mass size relation plotted in Fig.  \ref{fig:masstracks}, which is the scatter simply produced by the uncertainty in measuring the stellar mass, is $\sigma_{R_{1,mass}}$=0.047 dex. That is the contribution of the mass uncertainty to the total observed scatter of the $R_{\rm 1}$-mass size relation (i.e. $\sigma_{R_1}$=0.089 dex).

\section{Comparing the $R_{\rm 1}$-stellar mass relation with the  H{\sc i} size -mass relation of galaxies}
\label{app:hi}

An extremely tight scaling relation for galaxies is the one which links their total H{\sc i} mass with the diameter of their H{\sc i} discs defined by the location of the gas surface density  at 1 $M_{\rm \odot}$ pc$^{-2}$ \citep[][]{1997A&A...324..877B}. This relation spans over five orders of magnitude in H{\sc i} mass. A subsequent analysis of this relation has shown that its scatter  is extremely low: 0.06 dex and the slope almost exactly 1/2, i.e. 0.506$\pm$0.003 \citep{2016MNRAS.460.2143W}.  These values are suggestive of a uniform characteristic H{\sc i} surface density of $\sim$5 $M_{\rm \odot}$ pc$^{-2}$. 

In the case of the H{\sc i} size -mass relation of galaxies, the use of 1 $M_{\rm \odot}$ pc$^{-2}$ is a subjective choice and, contrary to the present work, there is no a priory physical motivation to select this value. Interestingly though, the scatter of the H{\sc i} size -mass relation is minimised when the H{\sc i} size is measured at surface densities between 1 to 2 $M_{\rm \odot}$ pc$^{-2}$. 

The reason  we cite this relation here is because of its potential connection to the size-mass relation we are exploring in this paper. In galaxy formation models \citep[see e.g.][]{2011MNRAS.418.1649L}, the H{\sc i} surface density is regulated by the conversion process from atomic to molecular hydrogen, i.e. H{\sc i}-to-H$_2$. Observationally, it is found that the H{\sc i} surface density saturates at $\sim$9 $M_{\rm \odot}$ pc$^{-2}$, and gas at higher surface densities are converted to molecular gas \citep[][]{2008AJ....136.2846B}. As H$_2$ and star formation are closely linked \citep[][]{2011ApJ...731...25K}, the H{\sc i}-to-H$_2$ process should be reflected in the star formation activity and, therefore, in the location of the gas density threshold for star formation in galaxies. As  galaxies observationally have a very similar average H{\sc i} surface density of $\sim$5 $M_{\rm \odot}$ pc$^{-2}$ (as reflected by the tight H{\sc i} size - mass relation of  galaxies), it seems reasonable to assume that the gas density threshold for  star formation  is also similar among very different galaxies. For this reason, we speculate that the small intrinsic scatter of the stellar mass - size relation we have explored in this work (based on the location of the gas density threshold for star formation) could be connected with or be a reflection of the tight  H{\sc i} size - mass relation of  galaxies.

\bsp	
\label{lastpage}
\end{document}